\documentclass[journal]{IEEEtran}
\usepackage{graphicx}
\usepackage{cite}
\usepackage{picinpar}
\usepackage{amsmath}
\usepackage{url}
\usepackage{flushend}
\usepackage[latin1]{inputenc}
\usepackage{colortbl}
\usepackage{soul}
\usepackage{multirow}
\usepackage{pifont}
\usepackage{color}
\usepackage{alltt}
\usepackage[hidelinks]{hyperref}
\usepackage{enumerate}
\usepackage{siunitx}
\usepackage{breakurl}
\usepackage{epstopdf}
\usepackage{pbox}

\usepackage[]{algorithm2e}
\usepackage{comment}

\ifCLASSINFOpdf
\else
\fi

\begin{document}
%




\title{Context-Aware Collaborative-Intelligence with Spatio-Temporal In-Sensor-Analytics\\ in a Large-Area IoT Testbed}

%

\author{Baibhab~Chatterjee,
        Dong-Hyun~Seo,
        Shramana~Chakraborty,
        Shitij~Avlani,
        Xiaofan~Jiang,
        Heng~Zhang,
        Mustafa~Abdallah,~\IEEEmembership{Student Member,~IEEE,}
        Nithin~Raghunathan,
        Charilaos~Mousoulis,~\IEEEmembership{Member,~IEEE,}
        Ali~Shakouri,~\IEEEmembership{Fellow,~IEEE,}
        Saurabh~Bagchi,~\IEEEmembership{Senior~Member,~IEEE,}
        Dimitrios~Peroulis,~\IEEEmembership{Fellow,~IEEE}
        and~Shreyas~Sen,~\IEEEmembership{Senior~Member,~IEEE}
\thanks{Manuscript received Month xx, 2xxx; revised Month xx, xxxx; accepted Month x, xxxx.
This work was supported in part by the SMART Films Consortium, Purdue University.}
\thanks{The authors are with the School of Electrical and Computer Engineering (ECE), Purdue University, West Lafayette, IN 47907 USA (e-mail: \{bchatte, seo60, chakra37, avlani, jiang175, zhan2614, abdalla0, nithin, cmousoul, shakouri, sbagchi, dperouli, shreyas\}@purdue.edu).}
\thanks{Color versions of one or more of the figures in this paper are available online at http://ieeexplore.ieee.org.}
\thanks{Digital Object Identifier 10.xxxx/JIOT.202x.xxxxxxx}}


\maketitle

\begin{abstract}
Decades of continuous scaling has reduced the energy of unit computing to virtually zero, while energy-efficient communication has remained the primary bottleneck in achieving fully energy-autonomous IoT nodes. This paper presents and analyzes the trade-offs between the energies required for communication and computation in a wireless sensor network, deployed in a mesh architecture over a 2400-acre university campus, and is targeted towards multi-sensor measurement of temperature, humidity and water nitrate concentration for smart agriculture. Several scenarios involving In-Sensor-Analytics (ISA), Collaborative Intelligence (CI) and Context-Aware-Switching (CAS) of the cluster-head during CI has been considered. A real-time co-optimization algorithm has been developed for minimizing the energy consumption in the network, hence maximizing the overall battery lifetime of individual nodes. Measurement results show that the proposed ISA consumes $\approx$467X lower energy as compared to traditional Bluetooth Low Energy (BLE) communication, and $\approx$69,500X lower energy as compared with Long Range (LoRa) communication. When the ISA is implemented in conjunction with LoRa, the lifetime of the node increases from a mere 4.3 hours to 66.6 days with a 230 mAh coin cell battery, while preserving more than 98\% of the total information. The CI and CAS algorithms help in extending the worst-case node lifetime by an additional 50\%, thereby exhibiting an overall network lifetime of $\approx$104 days, which is $>$90\% of the theoretical limits as posed by the leakage currents present in the  system, while effectively transferring information sampled every second. A web-based monitoring system was developed to archive the measured data in a continuous manner, and to report anomalies in the measured data.
\end{abstract}


\IEEEpeerreviewmaketitle


\section{Introduction}
\label{intro}

\IEEEPARstart{I}{nternet} of Things (IoT) is fast becoming one of the essential components of everyday life, through a plethora of devices available for smart homes, smart cities, wearable and implantable systems for healthcare, vehicle to everything (V2X) technologies and multiple other forms of wireless sensor networks (WSN). The use of these devices is expected to be so pervasive in future that CISCO predicts machine-to-machine (M2M) communication of 14.2 billion connected devices, just for IoT by 2022 \cite{VNI}. Due to limitations in size and available energy resources, these devices are required to perform sensing, computation and communication in the most energy-efficient manner that increases (1) the battery lifetime of the individual sensor nodes, as well as (2) the overall lifetime of the wireless network \cite{CompvsComm}. Furthermore, integration of multiple sensors on the same node helps in reducing the overall system-level energy and reduces the number of nodes to be deployed. However, it adds a significant amount of design complexity to individual nodes \cite{smart_city}.

\begin{figure}[!t]
\centering
\includegraphics[width=3.5in]{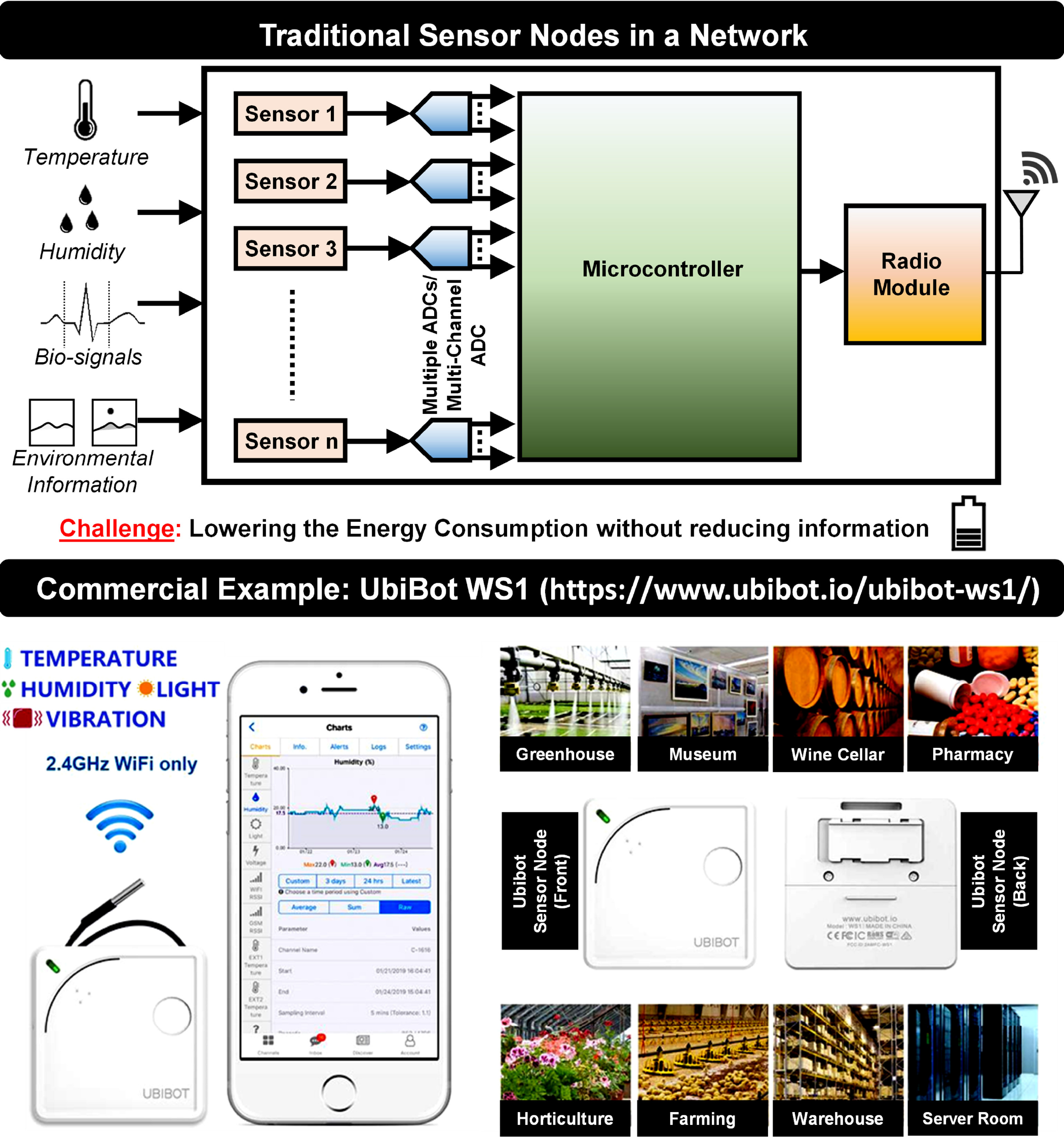}
\caption{Traditional Sensor Node Implementation: Multiple sensors are integrated using multiple ADCs/multi-channel ADCs with a microcontroller. A radio module is used to transmit the digital data through standard wireless techniques. A commercial example from UbiBot is shown as a reference \cite{UbiBot}.}
\label{smart_app}
\end{figure}

\begin{figure*}[!t]
\centering
\includegraphics[width=6.8in]{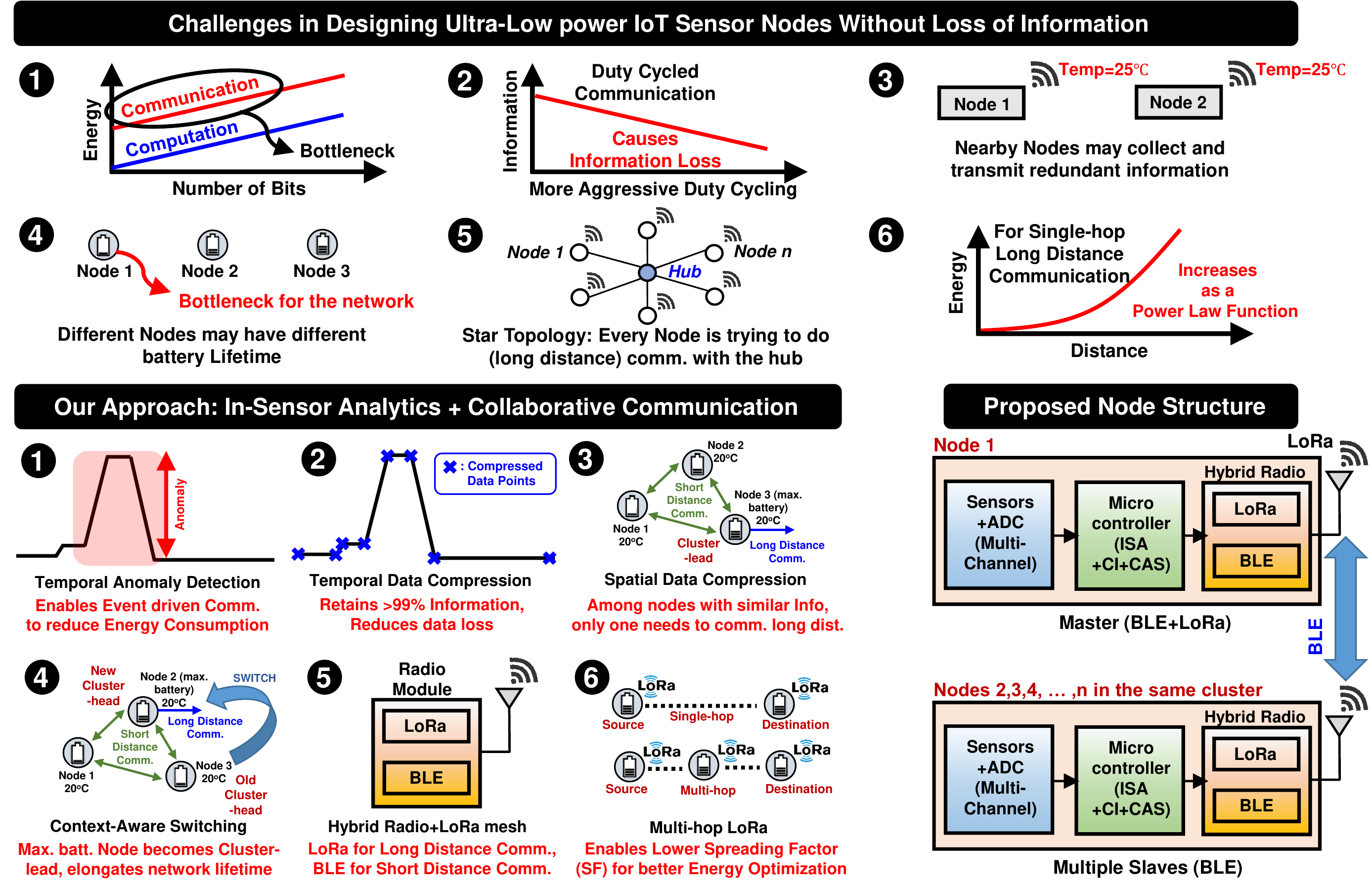}
\caption{The six challenges identified in designing ultra-low power IoT sensor node and our proposed solution, addressing all six challenges. The salient features of the implemented sensor node with In-Sensor Analytics (ISA) and Collaborative Intelligence (CI) are also described in brief: 1. Temporal Anomaly Detection, 2. Temporal Data Compression, 3. Spatial Data Compression/CI, 4. Context-Aware switching (CAS), 5. Hybrid Radio, 6. Multi-hop LoRa }
\label{contri}
\end{figure*}

\subsection{Background and Motivation}
A traditional multi-sensor node implementation for applications such as smart home or smart agriculture is shown in Fig. \ref{smart_app}. A number of sensors (for acquiring temperature, humidity, light, vibration or bio-signals) are connected to one node, the output of which is digitized using multiple analog-to-digital converters (ADCs), or multi-channel ADCs. The digitized signals are processed by a microcontroller and are sent out using a radio module to the cloud/remote servers where further analysis is performed on the data. A commercial multi-sensor example from UbiBot \cite{UbiBot} is also shown in Fig. \ref{smart_app} for Smart Agriculture \cite{smart_ag} that involves multiple sensors (temperature, humidity, water nitrate concentration etc.) and geo-positioning systems to monitor large farmlands in real-time and detects anomalous situations. UbiBot also provides software-based post-processing at the servers that helps the farmers to come up with optimum strategies as per the environmental conditions. In essence, such a system with a large-area WSN would require (1) integration of multiple sensors on the same node, (2) ultra-low energy consumption to increase network lifetime that avoids changing of the batteries through manual intervention, and (3) large area of coverage (normally in the range of 100's of meters to a few kilometers).

Fig. \ref{contri} identifies six challenges in designing a large-scale network of sensor nodes. The fundamental constraints on the battery lifetime primarily arises from the energy required for communication, which is usually orders of magnitude higher than computation energies (challenge 1). Traditionally, sensor nodes are duty cycled (i.e. they stay on for only a short time duration, and then put in sleep mode for a longer duration) to reduce the average power consumption of communication. However, in absence of a large amount of storage, duty cycling may also results in significant loss of data as the sensor node is in sleep mode for most of the time (challenge 2). Additionally, there could be challenges related to transmission of redundant/repeated data from multiple sensors due to their spatial proximity (challenge 3), reduction of overall network lifetime due to the node with the lowest battery remaining in a mesh topology (challenge 4), fast draining of batteries due to all sensor nodes trying to communicate with the hub in a star topology (challenge 5) and the power-law based steep rise in energy consumption for single-hop long distance communication performed by the sensor nodes (challenge 6).

\subsection{Proposed Solution}
In this work, we propose optimum strategies to reduce the energy consumption in a large-area WSN, through (1) in-sensor analytics (ISA: helps solving challenges 1-2) that enables event-driven communication of temporally compressed data without losing more than 2\% of information, (2) collaborative intelligence (CI: helps solving challenge 3) that helps in spatial data compression to lower the amount of redundancy in the transmitted data, and (3) context-aware switching (CAS: helps solving challenge 4) of the cluster-head during CI for extended network lifetime. To solve challenges 5-6, our implemented solution also features a hybrid radio module with a multi-hop LoRa mesh network for long range communication, and BLE for short range CI that achieves optimum communication energies. LoRa \cite{LORArev,whyLoRa} has been chosen over Sigfox or NB-IoT for long range communication because it allows development of private networks away from the licensed LTE frequency bands (LoRa uses the unlicensed 915 MHz ISM band in North America), has strong interference immunity along with encryption features and provides sufficient range (2-10 km). On the other hand, BLE was chosen over ANT or Zigbee for short range communication because of its simple pairing and link establishment options \cite{BLErev,BLESpec}.

Fig. \ref{contri} shows the salient features of our implemented multi-sensor IoT node. Temperature, humidity and water nitrate concentration can be sensed using the node for demonstration purpose. The read-out quantities are digitized and compressed in the temporal domain. Whenever the sensed quantity varies by an amount that is more than a predefined threshold, it is detected as an anomaly event. As soon as an anomaly occurs, the nodes communicate among themselves using BLE within a limited range ($\approx$10 m) to figure out if nodes within a close spatial proximity have collected similar data and anomaly patterns. We are calling this collaborative intelligence (CI) in context of this paper. The nodes with similar data pattern would then form clusters, and only one node in that cluster (that has the highest battery available) would communicate to a remote station using LoRa. This node would be called the cluster-head for the rest of the paper. Since LoRa consumes a high amount of energy, the role of the cluster-head would dynamically keep on switching to the node with the highest remaining battery life, through our context-aware switching (CAS) algorithm. As would be shown in section \ref{multi_hop}, a multi-hop architecture for LoRa proves to be better suited for a lower ratio of energy consumed/range covered.

 

\subsection{Related Work}
The energy to communicate one bit is usually orders of magnitude higher than the energy to compute on one bit \cite{WSN_high_comm_AKY, Barr}. For example, BLE communication usually requires 1 nJ/bit\cite{SenDAC}, LoRa communication requires $\approx$20 uJ/bit for a range of $>$ 1 km \cite{LORArev}, while computation in a 55 nm technology node costs $<$ 2 pJ/6-bit multiply and accumulate (MAC) operation \cite{AnveshaJSSC}. In \cite{Barr}, it was experimentally shown that the ratio of energy required to transmit one bit is $\approx 480-1270$ times higher than that of a 32-bit addition under varying channel conditions. This has motivated moving the burden of analysis from the cloud/servers to the sensor node itself, either partially or completely, thereby reducing the amount of communication required from the sensor node to the cloud. This has given rise to the paradigm of `Edge Intelligence' (EI) \cite{edge1, edge2}, which incorporates processing and/or machine learning capabilities at the sensor node. Analysis of data at the point of collection is particularly beneficial in IoT scenarios where energy is scarce or supply is intermittent. In \cite{ec_eg1}, Intel showed an IoT Device on a 14-nm node for a vision system used in agricultural platforms, and implemented multiple energy-scavenging techniques including duty cycling to reduce the overall average power. The SoC operated at 80 $\mu$W power for 0.4 V supply and 200 kHz clock frequency. However, duty cycling reduces the overall on-time and may result in significant amount of data loss. \cite{self_opt} optimizes the overall system energy (computation + communication) and bit error rate (BER) in a video sensor node for human detection and surveillance applications, by employing 4 modes of increasing computational complexity in the sensor node. These 4 modes involve (1) raw data transmission (R), (2) background subtraction followed by data communication (B), (3) feature extraction and communication (F), and (4) full classification and communication (C), for different path losses (40 dB-70 dB), classification algorithms and computation depths. For high path loss conditions (70 dB), the overall system energy is shown to reduce by factor of ~4X with respect to the case with raw data transmission. Intermittent computing and communication has been explored in \cite{edge2} which is necessary for extremely resource-constrained applications but requires embedded non-volatile memories (eNVM) and complex algorithms to store data before going into sleep mode and retrieve them when the device powers on again. For extremely resource-constrained nodes, simpler computation algorithms (such as data compression and anomaly detection for event-driven communication) is preferred \cite{CompvsComm}. Traditional data compression algorithms such as LZO or BWT requires high amount of memory ($\approx$10-100 kB), which makes them infeasible for many IoT applications \cite{CompvsComm}. Lightweight compression techniques such as miniLZO and S-LZW-MC \cite{sadler}, Principal Component Analysis for dimensionality reduction \cite{PCA} and frame difference based compression \cite{anvesha, SB_TOSN} have shown better applicability for extremely resource-constrained scenarios (such as large area smart agriculture). Furthermore, keeping in mind that the distributed IoT nodes for such applications may sometimes have similar sensor readouts due to spatial proximity, it might be more energy-efficient if a single node (out of the $n$ nodes having similar sensor readouts) communicate with the remote server. This notion has been exploited by some prior works in wireless sensor networks such as \cite{SB_TIBFIT, SB_Book}. However, to the best of our knowledge, there is no literature available on system optimization encompassing physical layer to the network layer, involving event-driven communication, temporal and spatial data compression and collaboration between multiple nodes using hybrid modes of communication, all of which are part of this paper.

\subsection{Our Contribution and Structure of this Paper}
In essence, the contributions of this paper are the following:
\begin{enumerate}[1)]
	\item Demonstration of anomaly detection and temporal data compression as algorithms for In-Sensor Analytics (ISA), which allows reduction of energy consumption due to communication in the sensor node without losing information.
	\item Demonstration of Collaborative Intelligence (CI) for spatial data compression that reduces data redundancy from multiple nodes which are in close geographical proximity.
	\item Demonstration of Context-Aware Switching (CAS) of cluster-head during CI, based on remaining battery life, to increase overall network lifetime.
	\item Development of a large-area multi-sensor network spanned over 2400 acres for demonstrating simultaneous physical layer and network layer energy optimization strategies including ISA, CI, CAS and multi-hop LoRa.
	\item Development of a web-based platform to continuously monitor $>$ 25 nodes for temperature, humidity, and water nitrate concentration. 
\end{enumerate}

The rest of the paper is organized as follows: We introduce and motivate the use of ISA, CI and CAS in Section \ref{theory} by building mathematical formulation for each concept. Section \ref{platform} describes the hardware platform on which the measurements are performed. Section \ref{implementation} presents the implementation details for the different strategies presented in this work, while Section \ref{measured} includes system-level measurement results. Section \ref{conclusion} summarizes our contributions and concludes the paper. 

\section{Analysis of the Problem Space}
\label{theory}
During the last decade, sensor technology and IoT has witnessed an unprecedented growth, and the number of nodes connected to back-end cloud servers have increased exponentially. This has resulted in a large amount of communication payload for the network, while the continuous demand for more information has increased the communication burden for individual sensor nodes \cite{Ningyuan-IMS}. Both of these aspects motivate the need for in-sensor analytics at the location of data collection, which envisions to reduce the overall communication power by enabling data compression and context-aware event-driven communication \cite{CompvsComm}.

\subsection{Theoretical Limits: Trading-off Communication Energy with Computation Energy at the System level}

Assuming unit energies (energy per bit) for computation and communication to be $E_{comp,u}$ and $E_{comm,u}$, respectively, the overall computation and communication energies ($E_{comp}$ and $E_{comm}$, respectively) in a system can be written as 

\begin{equation}
\begin{aligned}
	E_{comp} &= \left(E_{comp,u}\right) \times \text{Number of bits switched}\\
	E_{comm} &= \left(E_{comm,u}\right) \times \text{Number of bits transmitted}
\end{aligned}
\label{com_eqn}
\end{equation}

\begin{figure}[!t]
\centering
\includegraphics[width=3.5in]{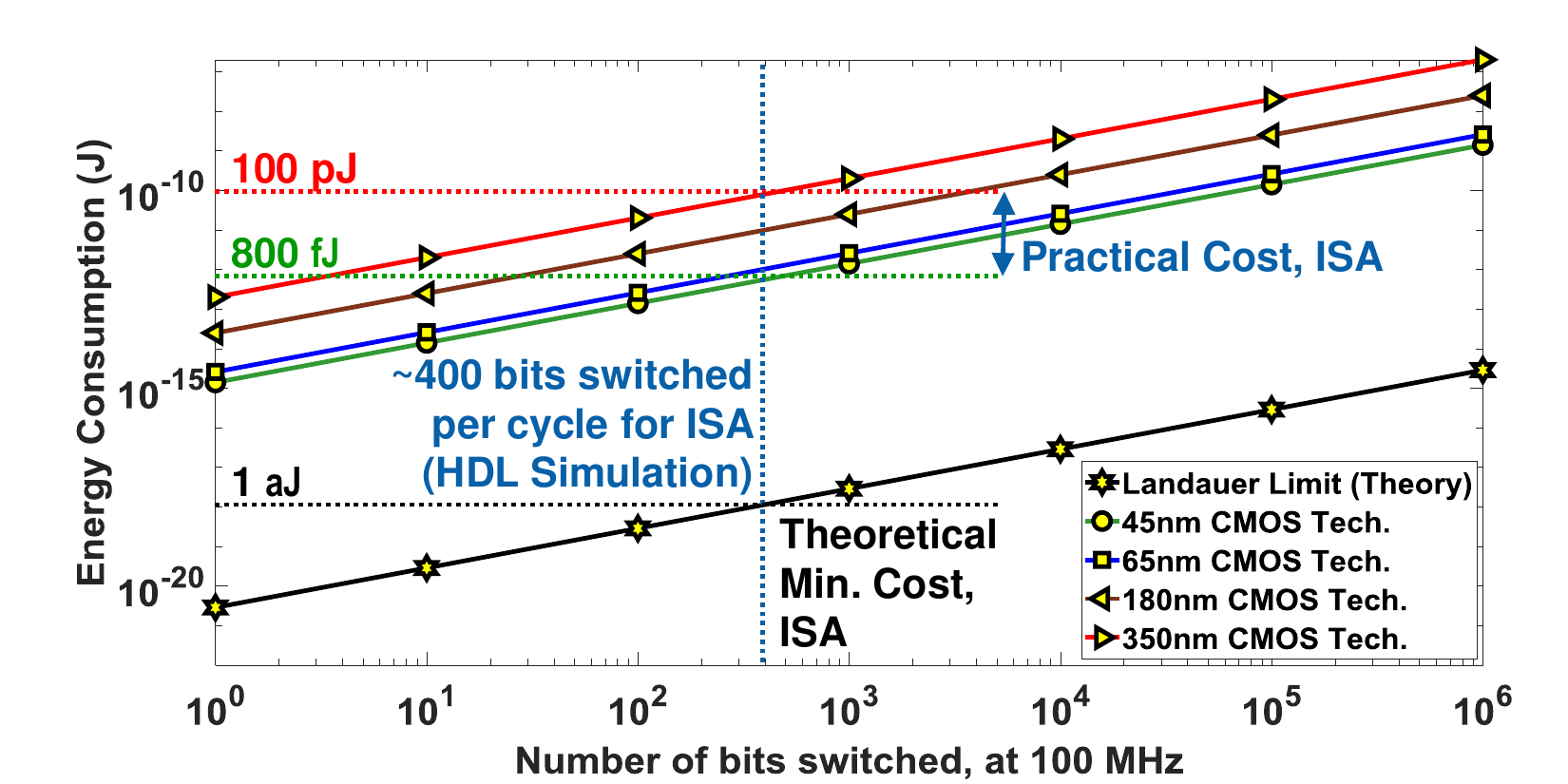}
\caption{Cost of computation for ISA (anomaly detection and data compression) in different process technologies, along with the theoretical Landauer limit. Leakage currents are neglected as the frequency of operation (100 MHz) is more than the frequencies where power is leakage-dominant \cite{NTV}.}
\label{comp_en}
\end{figure}

Since the computation energy primarily arises from  digital calculations, $\left(E_{comp,u}\right)$ can be approximated by $\left(E_{comp,u}\right) = CV^2$, which is the dynamic energy for a frequency of operation beyond the leakage-dominant region \cite{NTV}. For constant electric field scaling, both capacitance and voltage reduces with technology, and for an ideal technology that allows zero device capacitances, $\left(E_{comp,u}\right)$ reduces to the theoretical limits as posed by the Landauer's principle \cite{lndr}, and is shown in Eq. (\ref{lndr_eqn})

\begin{equation}
\begin{aligned}
	(E_{comp,u})_{th\_min} &= \kappa T \times \ln{2}
\end{aligned}
\label{lndr_eqn}
\end{equation}

where $\kappa$ is Boltzmann constant and $T$ is the absolute temperature. At room temperatures ($T$=298K),  $(E_{comp,u})_{th\_min}$ can be calculated to be $2.85 \times 10^{-21}$ J/bit. Fig. \ref{comp_en} shows the trend in computation energy obtained from HDL simulations for standard CMOS process technologies (such as $45$ nm, $65$ nm, $180$ nm and $350$ nm), with increasing number of bit-switching at a frequency of 100MHz (close to the frequency of operation of the nRF52 microcontroller in our implementation). For an HDL implementation of the proposed ISA (anomaly detection and data compression),  the amount of average energy consumption was found to be $\approx$800 fJ (80 $\mu$W power consumption at 100 MHz) using Synopsys Power Compiler in a standard $45$ nm CMOS technology. Using this number, we can estimate the average number of bits switching at the clock edge to be about 400 from Fig. \ref{comp_en} (which shows that the computation energy increases linearly at a rate of $\approx$2 fJ/bit in 45 nm technology).

On the other hand, if someday we are able to create a fictitious technology that offers zero capacitances, and allows us to build a zero-power receiver for communication, along with 100\% efficiency in the transmitter (Tx), $E_{comm,u}$ would eventually be limited by the free-space path loss ($FSPL$) of the physical channel, as the Tx still needs to transmit a power level which needs to be more than the receiver (Rx) sensitivity after considering the channel loss.  $FSPL$ is usually calculated using Frii's equation \cite{Friis_eqn} \cite{Friis_link_analysis}, and is shown in Eq. (\ref{friis_eqn}).

\begin{equation}
\begin{aligned}
	FSPL &= G_{Tx} . G_{Rx}(\frac{\lambda}{4{\pi}d})^{n}
\end{aligned}
\label{friis_eqn}
\end{equation}

where $G_{Tx}$ and $G_{Rx}$ are the transmitting and receiving antenna gains, respectively; $\lambda$ is the wavelength, $d$ is the distance between the Tx and the Rx, and $n$ is an empirical parameter that represents the fading margin (typically between 2 to 3). For a typical BLE protocol operating at $2.45$ GHz with $d=10$m, $FSPL$ can be optimistically estimated as $57$ dB ($n=2,G_{Tx}=2\text{ dB, } G_{Rx}=2 \text{ dB}$). If a state-of-the-art Rx with a sensitivity of $-100$ dBm \cite{salazar} is used in the system, then the Tx needs to transmit a minimum of $-43$ dBm, which directly translates to a power consumption of $50.119$ nW as a theoretical limit (since we assume zero capacitances and 100\% efficiency at the Tx).  With a typical data rate ($DR$) of 1 Mbps for BLE, the minimum theoretical energy efficiency for communication would then be $(E_{comm,u})_{th\_min} = 50.119 \times 10^{-15}$ J/bit, which is $> 10^7$ times higher than $(E_{comp,u})_{th\_min}$, as given by the Landauer principle.

The theoretical minimum energy/bit for communication is thus limited by the physical loss in the channel, and can be shown as Eq. (\ref{comm_eqn}).

\begin{equation}
\begin{aligned}
	(E_{comm,u})_{th\_min} &= \frac{Rx_{sen}}{FSPL \times \eta . DR}
\end{aligned}
\label{comm_eqn}
\end{equation}

where $Rx_{sen}= \kappa T \times NF \times SNR \times BW$ is the receiver sensitivity (assuming conjugate matching \cite{razavi_RF} with $NF$= Noise Figure, $SNR$ = Signal to Noise Ratio required for a particular modulation, $BW$ = Rx bandwidth), $\eta$ is Tx's efficiency and $DR$ is the communication data rate.

\begin{figure}[!t]
\centering
\includegraphics[width=3.5in]{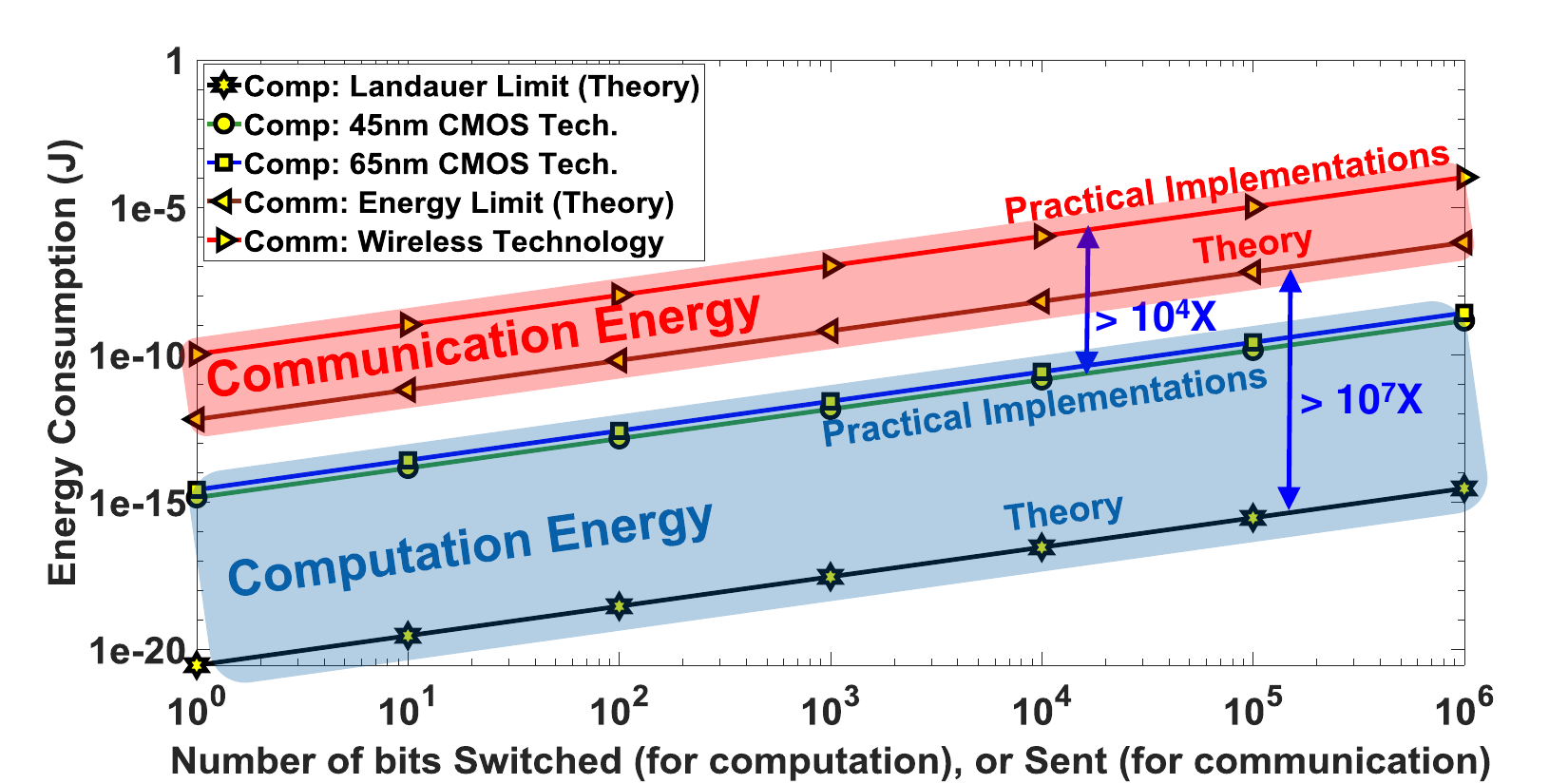}
\caption{Comparison of Communication and Computation energies \cite{CompvsComm} (both theoretical and from standard implementations \cite{30pJb} that show that Communication energy is $>10^4$ times of Computation energy ( with same number of bits). Leakage is again ignored in the analysis as in Fig. \ref{comp_en}. This motivates the need for event-driven communication.}
\label{comm_en}
\end{figure}

Fig. \ref{comm_en} summarizes the foregoing discussions and compares $E_{comm}$ with $E_{comp}$ for the same number of bits transmitted, or switched \cite{CompvsComm}. Even with today's scaled technologies, state-of-the-art wireless transceivers \cite{30pJb} consume $>10^4$ times more energy than computational bit-switching in standard $45$ nm and $65$ nm nodes. This bottleneck-analysis signifies that there is a possibility of reducing the overall system power through ISA (selective data transmission with compression) which reduces $E_{comm}$ with additional $E_{comp}$. Please note that this conclusion will hold true as long as the ratio of number of bits switched during ISA and the reduction in the number of bits transmitted because of ISA is less than the practical ratios of $\left(E_{comm}/E_{comp}\right)$. The anomaly detection algorithm enables immediate data transmission only when an aberration from the expected pattern is observed, whereas the data compression helps is reconstructing valuable information even when data is sent only sparingly.


\subsection{Communication Energy Bottleneck: Payload and Switching}
Assuming that the sensor samples and communicates data every $N$ seconds, the total on-time for the communication modules over $n_{sec}$ seconds is $T_{on} = \frac{n_{bits} \times n_{sec}}{DR \times N}$, where $n_{bits}$ is the number of bits transmitted per sample. The total energy consumed during $n_{sec}$ seconds can then be written as shown in Eq. (\ref{N_eqn}).

\begin{equation}
\begin{aligned}
	E_{comm} &= T_{on}.I_{on} + T_{off}.I_{comp,lkg} + 2.T_{tran}.I_{on}.\frac{n_{sec}}{N}
\end{aligned}
\label{N_eqn}
\end{equation}

where $I_{on}$ is the current consumption when the communication modules are on (along with computation and leakage), $T_{off} = (n_{sec} - T_{on})$, $I_{comp,lkg}$ is the computation and leakage current consumption when the communication modules are off, and $T_{tran}$ is the transient power-on/switching time (or power-off time, hence the factor 2) added with set-up time for the communication modules. From Eq. (\ref{N_eqn}), it is evident that when $T_{on} \ll 2.T_{tran}.\frac{n_{sec}}{N}$ (i.e. when $\frac{n_{bits}}{DR} \ll 2.T_{tran}$), the communication is limited by the switching energy. On the other hand, when $2.T_{tran} \ll \frac{n_{bits}}{DR}$, communication is limited by the payload (i.e. number of bits). For all practical purposes, $2.T_{tran}$ is usually less than a few ms \cite{nRFSpec}, while the number of bits per sample is much much smaller than the data rate, making communication predominantly limited by the number/energy of switching.

\begin{figure}[!t]
\centering
\includegraphics[width=3.5in]{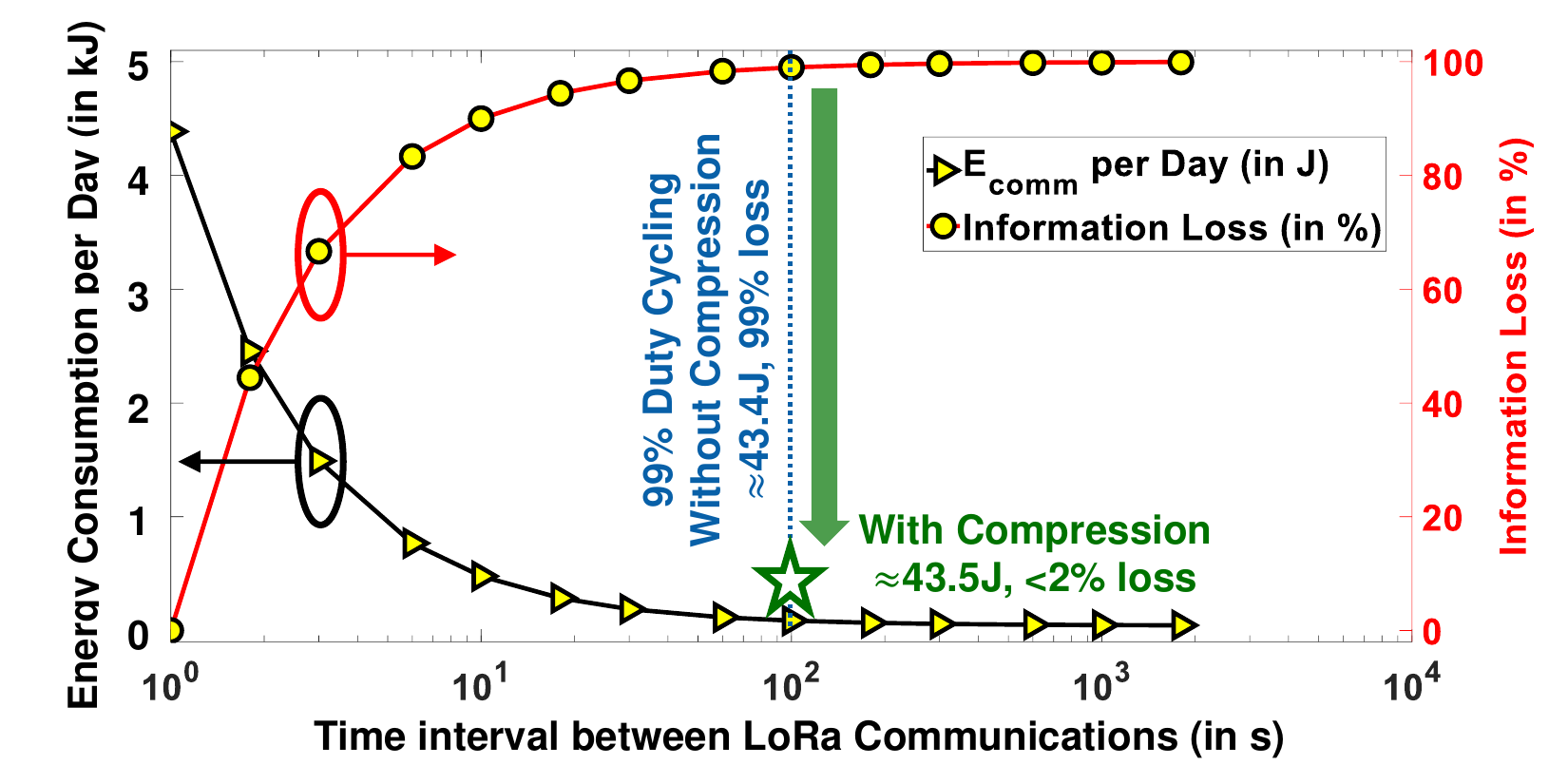}
\caption{Communication Energy and Information loss as a function of duty-cycling the Communication Module, motivating the need for Data Compression and storage for reducing energy consumption without loss of information.}
\label{duty_N}
\end{figure}

To reduce the amount of switching energy, previous literature (for example, \cite{ec_eg1}, \cite{ec_eg2}) preferred a duty cycle based approach where power consumption is minimized by increasing $N$. However, increasing $N$ also increases the probability of useful information being lost. This scenario is portrayed in Fig. \ref{duty_N} where the communication energy per day and information loss is plotted as a function of $N$ for measured values of $I_{on}$, $I_{comp,lkg}$, and $T_{tran}$ for LoRa communication in a Nordic nRF52 platform (to be discussed in Section \ref{platform}). Taking $N=1$ as a baseline, $N=100$ would result in $50$X reduction in energy with 99\% loss in information content. To avoid the information loss, all the data in the previous $N$ second interval needs to be stored and subsequently transmitted during communication. However, communication may now become a strong function of payload if $2.T_{tran} \ll \frac{n_{bits} \times N}{DR}$, warranting the need for a data compression algorithm to reduce the total number of bits to be stored and transmitted.

\subsection{Collaborative Intelligence: United, we stand?}
The concepts of anomaly detection and data compression discussed thus far helps in ``temporal compression'' of the sensor data. When several sensors are deployed in close geographical proximity, it is possible that multiple sensors would receive very similar data. In such a scenario, the energy consumption for iso-information would be sub-optimal if all the sensors transmit their data to a long-distance hub. Instead, the sensors can talk among themselves using a short-distance, low-power communication protocol such as BLE, form clusters based on the gathered data, and elect a cluster-head which will communicate a ``spatially compressed'' version of the data (non-redundant data obtained from multiple spatially separated sensors) to a long-distance receiver using a high-power protocol such as LoRa. The total energy savings in the network when $n(\geq 2)$ nodes can be clustered together is given by Eq. (\ref{ci_eqn}).

\begin{equation}
\begin{aligned}
	E_{savings,CI} &= \textit{Energy without CI - Energy with CI}\\
	               &= n \times E_{lr} - \left[n \times \left(E_{sr} + E_{comp,CI}\right) + E_{lr} \right]\\
	               &= (n-1)\times E_{lr} - n \times \left(E_{sr} + E_{comp,CI}\right)
\end{aligned}
\label{ci_eqn}
\end{equation}

\begin{figure}[!t]
\centering
\includegraphics[width=3.5in]{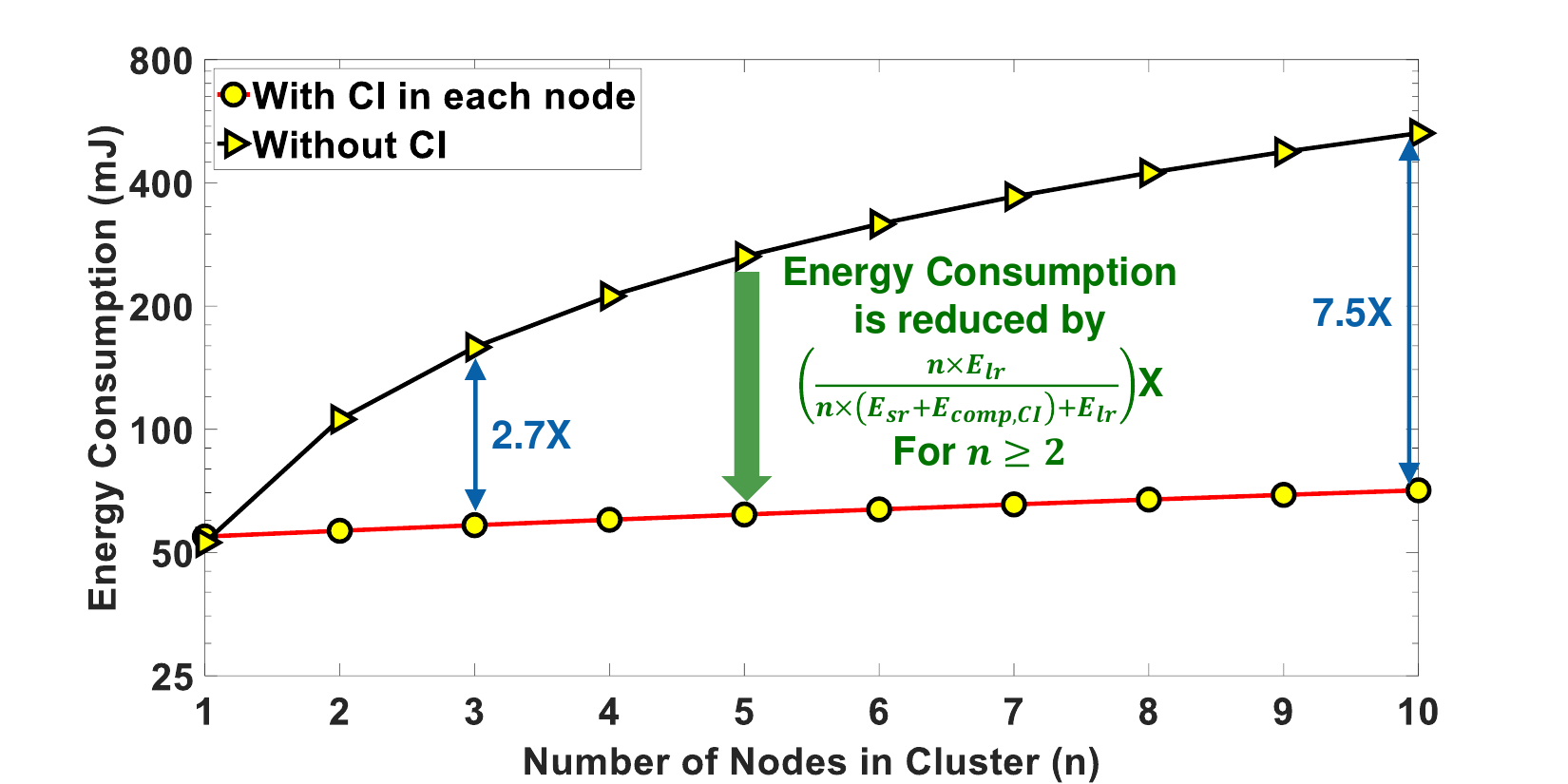}
\caption{CI as a method of reducing the cluster's energy consumption (during one communication cycle of 30 minutes) for $n \geq 2$. Leakage is ignored in this analysis.}
\label{ci}
\end{figure}

where $E_{lr}$ and $E_{sr}$ are the energies for long-range (LoRa) and short-range (BLE) communication, respectively, while $E_{comp,CI}$ is the energy for CI implemented in each node. For large $n$, $E_{savings,CI} \to n \times (E_{lr}-E_{sr}-E_{comp,CI})$, which means that the energy is reduced by a factor of  $\left(\frac{E_{lr}}{E_{sr}+E_{comp,CI}}\right)$ when CI is employed. Fig. \ref{ci} shows the reduction in power consumption with CI in the nRF52 platform with measured values of $E_{lr}$ ($\approx 50$ mJ), $E_{sr}$ ($=359$ $\mu$J) and $E_{comp,CI}$ ($=804$ nJ), with the assumption that communication happens once every $30$ minutes while computation happens once every second, making the effective $E_{comp,CI} = 1.4$ mJ in between two communication cycles.


\subsection{Context-Aware switching: Is change the only constant?}

\begin{figure}[!t]
\centering
\includegraphics[width=3.5in]{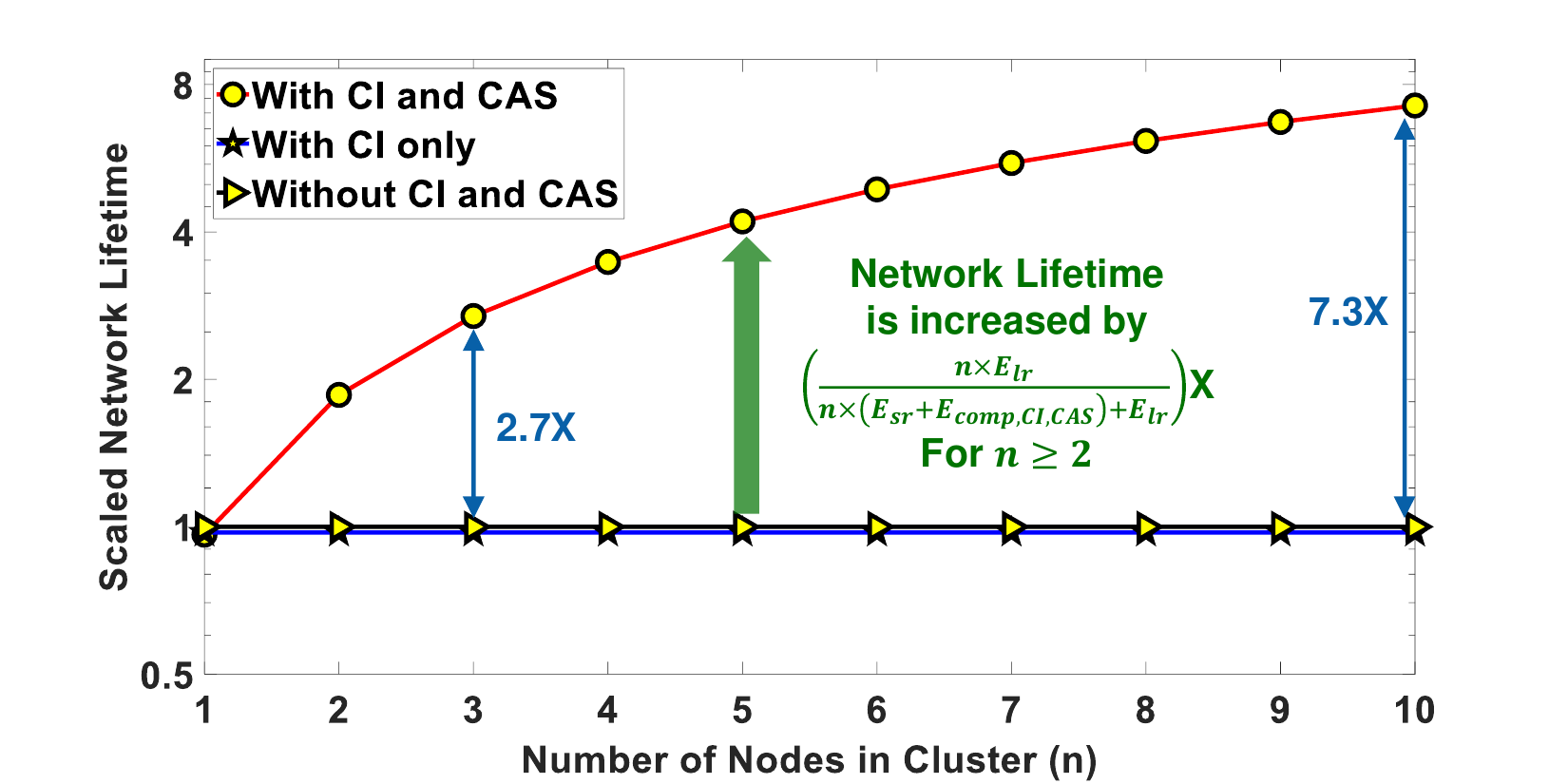}
\caption{CAS as a method of increasing network lifetime for $n \geq 2$. Leakage is ignored in this analysis.}
\label{ci_cas}
\end{figure}

Even though collaborative intelligence reduces the overall energy consumption in the network, it does not necessarily increase the network lifetime when all nodes are designed with similar resources. To understand this, let us consider the following cases:
\begin{enumerate}[1)]
	\item \emph{Case 1: CI with fixed cluster-head}\\
	For a stationary environment and immobile nodes, if the same cluster-head is chosen every time in an ad-hoc manner, the battery for the cluster-head will drain faster - indicating that the network lifetime will be limited by the battery-life of the cluster-head. Since all nodes in the network start with the same amount of resources (i.e. same sized batteries), CI in this scenario is detrimental to the overall network lifetime, which is given by Eq. (\ref{NL_eqn_case1}).
        \begin{equation}
        \begin{aligned}
	        NL_{CI} &= \frac{E_{BATT}}{\textit{energy consumed in bottleneck node (cluster-head)}}\\ 
	           &= \frac{E_{BATT}}{E_{sr} + E_{comp,CI} + E_{lr}}\\
        \end{aligned}
        \label{NL_eqn_case1}
        \end{equation}
        where $E_{BATT}$ is the battery capacity of each of the sensor nodes. Interestingly, network lifetime in this scenario will be even lower than the case with no CI and CAS $\left(\frac{E_{BATT}}{E_{lr}}\right)$.
	\item \emph{Case 2: CI and CAS with variable cluster-head}\\
	If the cluster-head is chosen based on the context (for example, instantaneous battery-life and/or distance from the receiver hub), network lifetime can be significantly improved, as the cluster-head will keep on switching till the point when all the nodes exhaust their batteries. The overall network lifetime in this scenario is presented in Eq. (\ref{NL_eqn_case2}).
	    \begin{equation}
        \begin{aligned}
	        NL_{CI,CAS} &= \frac{n \times E_{BATT}}{n \times \left(E_{sr} + E_{comp,CI,CAS}\right) + E_{lr}}\\
        \end{aligned}
        \label{NL_eqn_case2}
        \end{equation}
	where $E_{comp,CI,CAS}$ is the energy for CI and CAS implemented simultaneously in each node (916nJ during each sensing cycle of 1 second from measured results). Thus, context aware switching can increase the network lifetime by $\left(\frac{E_{lr}}{E_{sr}+E_{comp,CI,CAS}}\right)$ times when $n$ is large. This is presented in Fig. \ref{ci_cas} which shows the network lifetime as a function of $n$ in $3$ scenarios: (1) without CI and CAS, (2) with CI only, and (3) with simultaneous CI and CAS. The y-axis is scaled by the first scenario: network lifetime without CI and CAS. The lifetime reduces by 4\% when only CI is employed, but increases drastically when both CI and CAS are implemented for $n \geq 2$.

\end{enumerate}


\subsection{Multi-Hop LoRa: Distributing the Load}
\label{multi_hop}

In the limiting case, the sensitivity of the LoRa Rx should be equal to the transmitted power multiplied with $FSPL$ (which is a function of the distance $d$ between the Tx and Rx, according to eq. (\ref{friis_eqn})). However, it needs to be kept in mind that LoRa uses Chirp Spread Spectrum (CSS) with a spreading factor $SF$ (ranging from 7-12, which signifies that each symbol is spread through a chirp code whose length is in the range $2^7$-$2^{12}$), that improves the Rx sensitivity by a factor of $2^{SF}$. The Range ($d_{max}$) for LoRa communication was analyzed in \cite{LORArev}, and is shown in eq. (\ref{dmax_lora_eq}).

\begin{equation}
        \begin{aligned}
	        d_{max} &= \left(\left(\frac{\lambda}{4{\pi}}\right)^{n} \times \frac{P_{Tx} \times 2^{SF}}{\kappa T \times NF \times SNR \times BW}\right)^{1/n}\\ 
        \end{aligned}
        \label{dmax_lora_eq}
\end{equation}

where $P_{Tx}$ is the transmitted power and $SF$ is the spreading factor. All other terms are same as that defined in eq. (\ref{friis_eqn} and \ref{comm_eqn}). For $P_{Tx}$ = 7 dBm, $NF$ = 3.5 dB, $SNR$ = 15 dB, $BW$ = 125 kHz (LoRa standards as explained in \cite{LORArev}) and  $SF$ = 7, $d_{max} \approx$ 1.25 km, while  $d_{max} \approx$ 4 km for $SF$ = 12. For multiple hops, the range increases linearly.

The Energy consumption per bit for a packet of LoRa transmission can be written in the form as shown in eq (\ref{ebit_lora_eq}).

\begin{equation}
        \begin{aligned}
	        E_{bit,LoRa} &= \frac{P_{cons,Tx} \times Bytes_{Packet} \times {T_{Byte}}}{8 \times Bytes_{Payload}}\\ 
        \end{aligned}
        \label{ebit_lora_eq}
\end{equation}

where $P_{cons,Tx}$ is the power consumed in the Tx for a particular $P_{Tx}$ (for example, $P_{cons,Tx}$ = 95.4 mW for $P_{Tx}$ = 7 dBm and $P_{cons,Tx}$ = 412.5 mW for $P_{Tx}$ = 20 dBm \cite{LORArev}), $T_{Byte} = \frac{2^{SF}}{BW}$ is the total time to send one Byte through LoRa, $Bytes_{Payload}$ is the actual number of Payload Bytes in the packet and $Bytes_{Packet}$ is the total number of bytes in the packet (which, as shown in \cite{LORArev} is equal to $\left( 8+max\left(ceil\left( \frac{8Bytes_{Payload} - 4SF+16+28-20H}{SF-2DE} \right) \times \frac{1}{CR},0 \right) \right) \\ + Bytes_{Preamble} + 4.25$, with $CR$ being the code rate, $Bytes_{Preamble}$ is the number of Bytes in the Preamble, $H=1$ in presence of a header and $H=0$ in absence of it, $DE=1$ when low data rate optimization is enabled and $DE=0$ otherwise).

\begin{figure}[!t]
\centering
\includegraphics[width=3.5in]{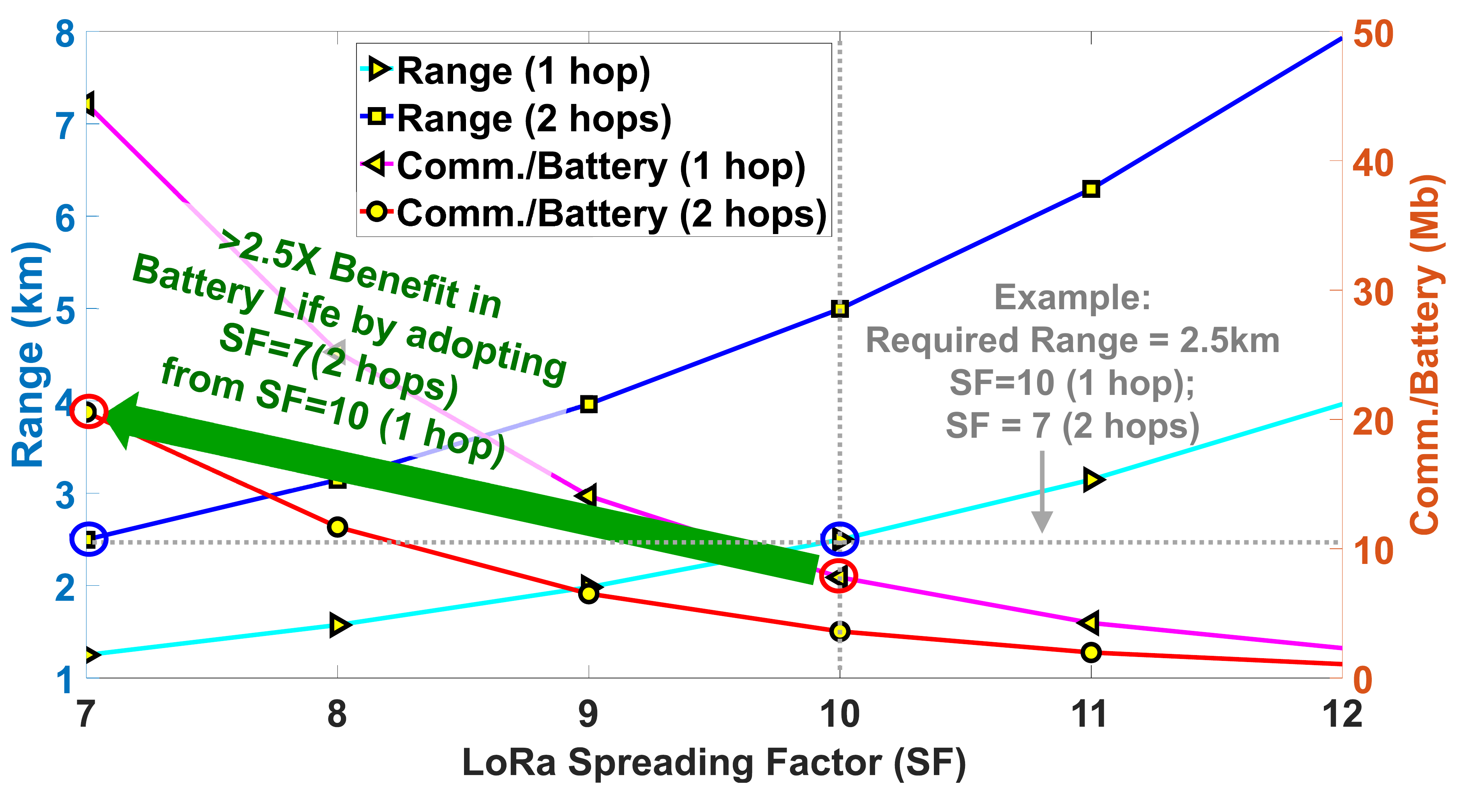}
\caption{Energy Savings and Improvement in Node Lifetime due to Multi-Hop LoRa as compared to Single-Hop LoRa. For a range of 2.5km, 2-hop LoRa requires $SF$ = 7, while 1-hop LoRa requires $SF$ = 10 ($P_{Tx}$ = 7dBm). By adopting the 2-hop architecture, the amount of communication (in Mb) with a single 230mAh battery can be increased by $>$ 2.5X due to lower $SF$.}
\label{SF_LoRa}
\end{figure}

For multiple hops ($n_{hops}$), we would incur the overhead of ($n_{hops}-1$) additional Tx and ($n_{hops}-1$) additional Rx devices, leading to a modified version of eq (\ref{ebit_lora_eq}), as presented in eq (\ref{ebit_lora_eq_nhop}).

\begin{equation}
        \begin{aligned}
	        E_{bit,LoRa,n\_{hops}} &= \frac{P_{cons,TRx,n\_hops} \times Bytes_{Packet} \times {T_{Byte}}}{8 \times Bytes_{Payload}}\\ 
        \end{aligned}
        \label{ebit_lora_eq_nhop}
\end{equation}

where $P_{cons,TRx,n\_hops}$ is the total power consumed in the $n_{hops}$ number of Tx and ($n_{hops}-1$) number of Rx devices, and is given by $\left( n_{hops} \times P_{cons,Tx} + (n_{hops}-1) \times P_{cons,Rx} \right)$. Fig. \ref{SF_LoRa} Shows the range on the left y-axis and the number of bits that can be communicated with a 230 mAh coin cell battery ($\frac{230 \times 10^{-3} \times 3600}{E_{bit,LoRa,n\_{hops}}}$) on the right y-axis as a function of $SF$, with $n_{hops}$ as a parameter (with $P_{Tx}$ = 7 dBm, $P_{cons,Tx}$ = 95.4 mW, $P_{cons,Rx}$ = 15.2 mW, $CR$ = 4/5,  $BW$ = 125 kHz, $H$ = 0, $DE$ = 0, $Bytes_{Preamble}$ = 8, $Bytes_{Payload}$ = 240, $NF$ = 3.5 dB and $SNR$ = 15 dB). As an example, if the required range for LoRa communication is 2.5 km, we would require $SF$ = 10 for $n_{hops}$ = 1  and $SF$ = 7 for $n_{hops}$ = 2. Since  $E_{bit,LoRa,n\_{hops}}$ is proportional to $2^{SF}$ and $Bytes_{Packet}$ (which is a weak function of $SF$ as shown in \cite{LORArev}), the benefit in $E_{bit,LoRa,n\_{hops}}$ (also the number of bits that can be communicated with a 230 mAh coin cell battery) can be represented as shown in eq. (\ref{benefit_nhop}).

\begin{equation}
        \begin{aligned}
	        \text{Benefit}_{n\_{hops}} &= \frac{2^{SF_B-SF_A} \times Bytes_{Packet_B} \times P_{cons,Tx}}{Bytes_{Packet_A} \times P_{cons,TRx,n\_hops}}\\ 
        \end{aligned}
        \label{benefit_nhop}
\end{equation}

where $SF_B$ is the $SF$ required for a particular range before opting for multi-hop and $SF_A$ is the $SF$ required for the same range after opting for multi-hop. $Bytes_{Packet_B}$ is the number of bytes in the LoRa packet before opting for multi-hop and $Bytes_{Packet_A}$ is the number of bytes in the LoRa packet after opting for multi-hop. For this particular example, $SF_B$ = 10, $SF_A$ = 7 and $n_{hops}$ = 2, leading to $Bytes_{Packet_B}$ = 249.25, $Bytes_{Packet_A}$ = 354.25 and $Benefit_{n\_{hops}} = \frac{2^3 \times 249.25 \times 95.4}{354.25 \times (2 \times 95.4 + 15.2)} \approx 2.6$X.


\section{Platform Description}
\label{platform}

\begin{figure}[!t]
\centering
\includegraphics[width=3.5in]{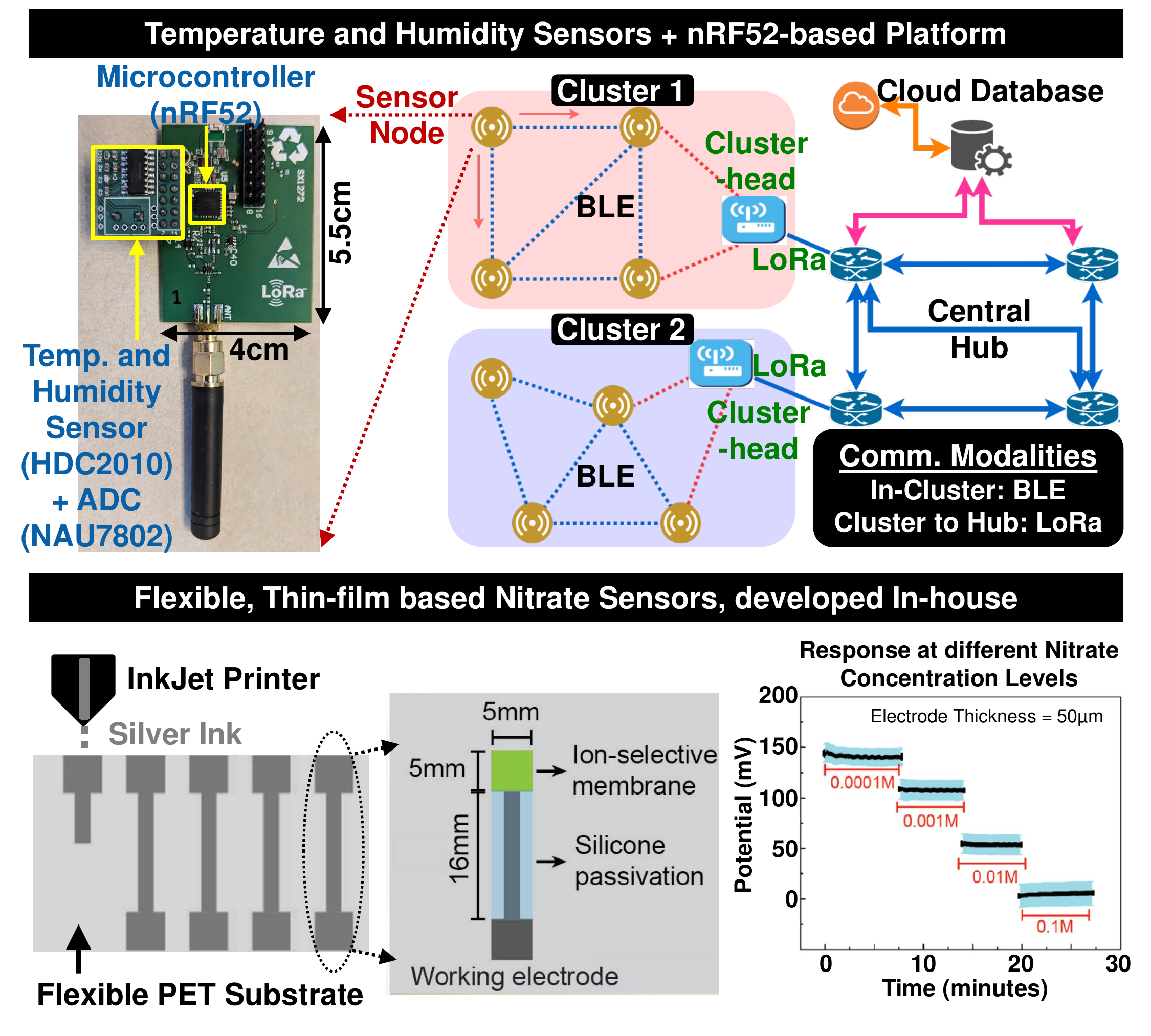}
\caption{Sensor Platform: nRF52-based sensor Node, 2 clusters as examples, and the WSN (with BLE for  local communication, LoRa for long-range communication) for implementation of ISA, CI and CAS. The thin-film based flexible nitrate sensors \cite{nitrate2}, developed in-house, are also shown along with their performance characterization curve.}
\label{platformfig}
\end{figure}

The prototype sensor (Fig. \ref{platformfig}) comprises a temperature and humidity Sensor (TI HDC2010), a long-range (LoRa) radio (SX1272), and a BLE-enabled embedded processor (nRF52) that implements ISA, CI and CAS. Sensor nodes within $\approx 300$ m$^2$ area can communicate among themselves using BLE for short-range communication. Whenever a temperature/humidity anomaly is detected, the node immediately sends that information to nearby sensors using BLE (as an input to CI algorithm). The data compression and CI algorithms compress (both temporally and spatially) the data while the cluster-head (elected using CAS) sends the compressed data to a LoRa receiver at a distance of $\approx$2 km. Both the BLE and the LoRa are designed as mesh networks which increases coverage and redundancies to reduce failure cases.

 A thin-film based, flexible, screen printed nitrate sensor is also developed and is fabricated in-house \cite{nitrate2} by printing silver ink onto a Polyethylene Terephthalate (PET) substrate, as shown in Fig. \ref{platformfig}. The nitrate concentration is represented by a potential difference between a working electrode and a reference electrode. The sensing area of the working electrode is coated with an ion-selective membrane (ISM), targeted towards nitrate ions, while the rest of the sensor is passivated with silicone. The reference electrodes are passivated in the sensing area as well. Further details on the nitrate sensor can be found in \cite{nitrate2}.


\section{Implementation Details}
\label{implementation}

\begin{figure}[!t]
\centering
\includegraphics[width=3in]{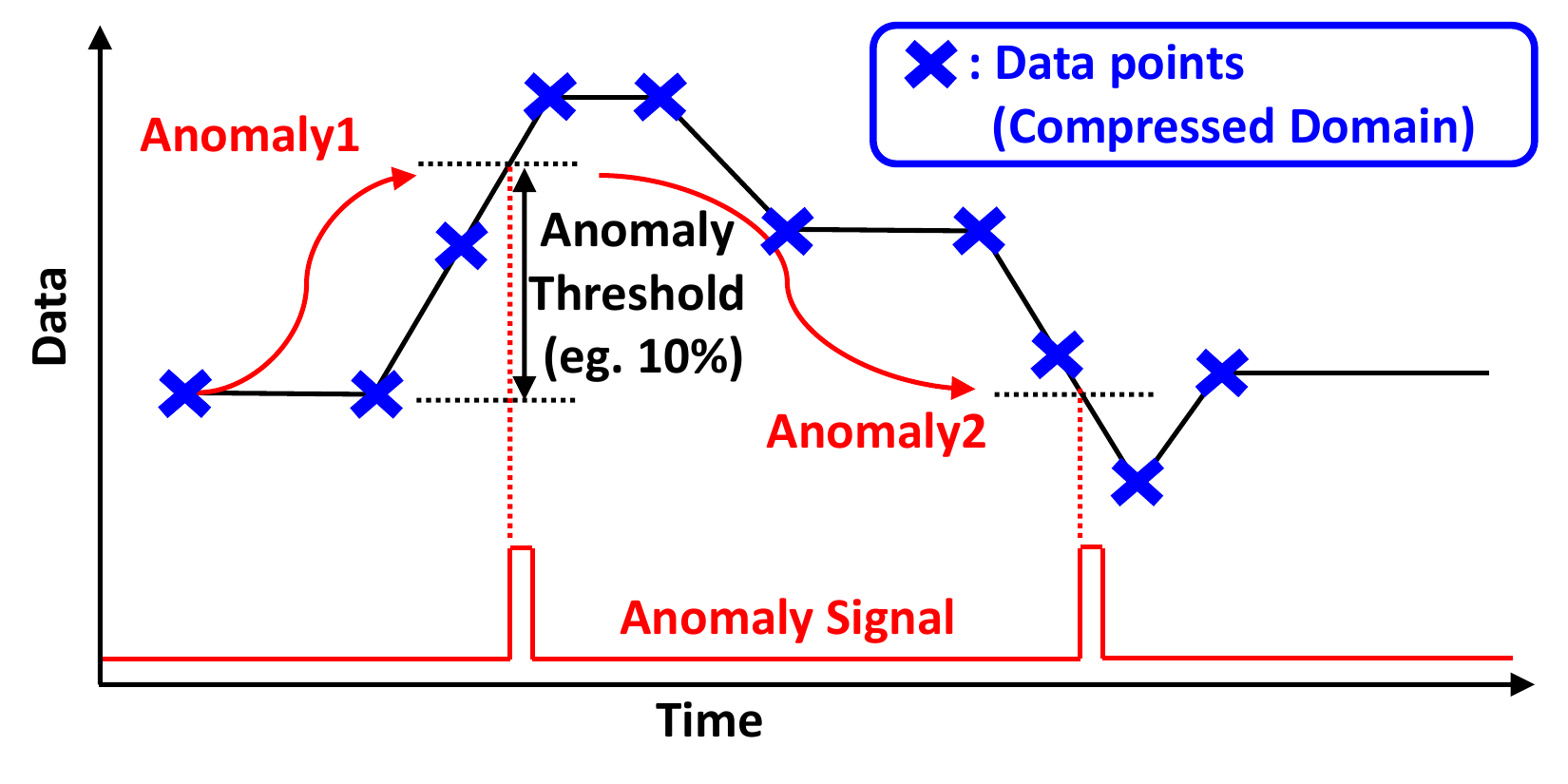}
\caption{Illustration of Anomaly Detection and Data Compression.}
\label{ad_dc}
\end{figure}

\subsection{ISA: Anomaly Detection and Data compression}
Fig. \ref{ad_dc} illustrates the basics of the lightweight anomaly detection and data compression algorithms, implemented in this work. Two predefined thresholds are utilized for each of the sensed quantities - one for anomaly detection ($x\%$ change in the data from the last anomaly) and the other for compressing the data ($y\%$ change in data from the last recorded data point). These thresholds, $x\%$ and $y\%$ are calculated offline by analyzing the sensed quantities over $>4$ weeks by using a k-means clustering algorithm, and a provision is kept to alter these thresholds.

\begin{algorithm}
\begin{small}
 \KwData{Continuous data stream-in from sensor/ADC}
 \KwResult{Anomaly Detection in Data}
 initialization\;
 \While{data is being sensed}{
  read threshold (\textit{x}) from k-means clustering algorithm\;
  \eIf{data differs by $>$ x\% from last reported anomaly (ISA)}{
   activate BLE communication\;
   broadcast anomaly information to nearby sensors (CI)\;
   deactivate BLE communication\;
   analyze received BLE data for CAS (form clusters and perform spatial compression)\;
        \eIf{Current Node has the highest battery life in the cluster}{
        activate LoRa communication\;
        send temporally and spatially compressed data stream to Hub\;
        deactivate LoRa communication\;
        }{
        wait for the next anomaly/passing of an hour\;
        stay in low-power sense and compute mode\;
        }
   }{
   wait for the next anomaly/passing of an hour\;
   stay in low-power sense and compute mode\;
  }
 }
 \caption{Anomaly Detection followed by BLE/LoRa communication}
\end{small}
\end{algorithm}

\begin{algorithm}
\begin{small}
 \KwData{Continuous data stream-in from sensor/ADC}
 \KwResult{Data Compression}
 initialization\;
 \While{data is being sensed}{
  read threshold (\textit{y}) from k-means clustering algorithm\;
  \eIf{data differs by $>$ y\% from last saved data-point}{
   save current data-point as latest in array\;
   }{
   discard data-point\;
  }
 }
 \caption{Temporal Data Compression Algorithm}
\end{small}
\end{algorithm}

\begin{figure}[!t]
\centering
\includegraphics[width=3.5in]{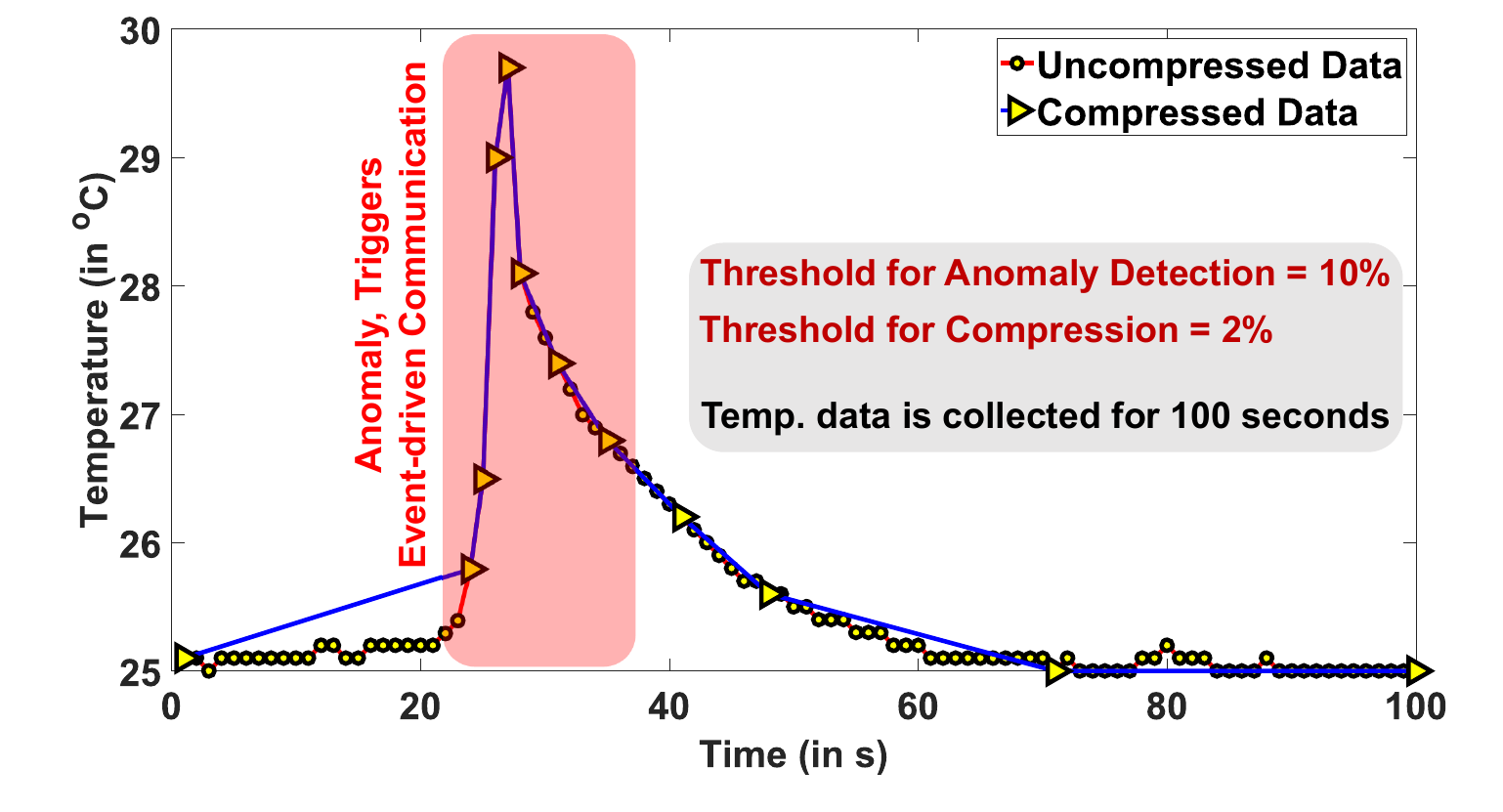}
\caption{An example of the measured temperature profile, with and without data compression. The threshold for compression is kept as 2\%.}
\label{compress1}
\end{figure}

An example of the measured temperature profile is shown in Fig. \ref{compress1}. The threshold for anomaly detection is kept at 10\%, while the threshold for data compression and record is kept at 2\%. Over 100 seconds, we collect 100 samples of the uncompressed data, while only 12 data points are saved through the temporal data compression algorithm (compression ratio = 100/12 = 8.33). Out of these 12 data points, 10 are because of a temperature anomaly created during 24-30 seconds, and due to the gradual change in data while the sensor comes back to normal temperatures. Without such an anomaly, the degree of compression would be even higher. 

\begin{figure}[!t]
\centering
\includegraphics[width=3.5in]{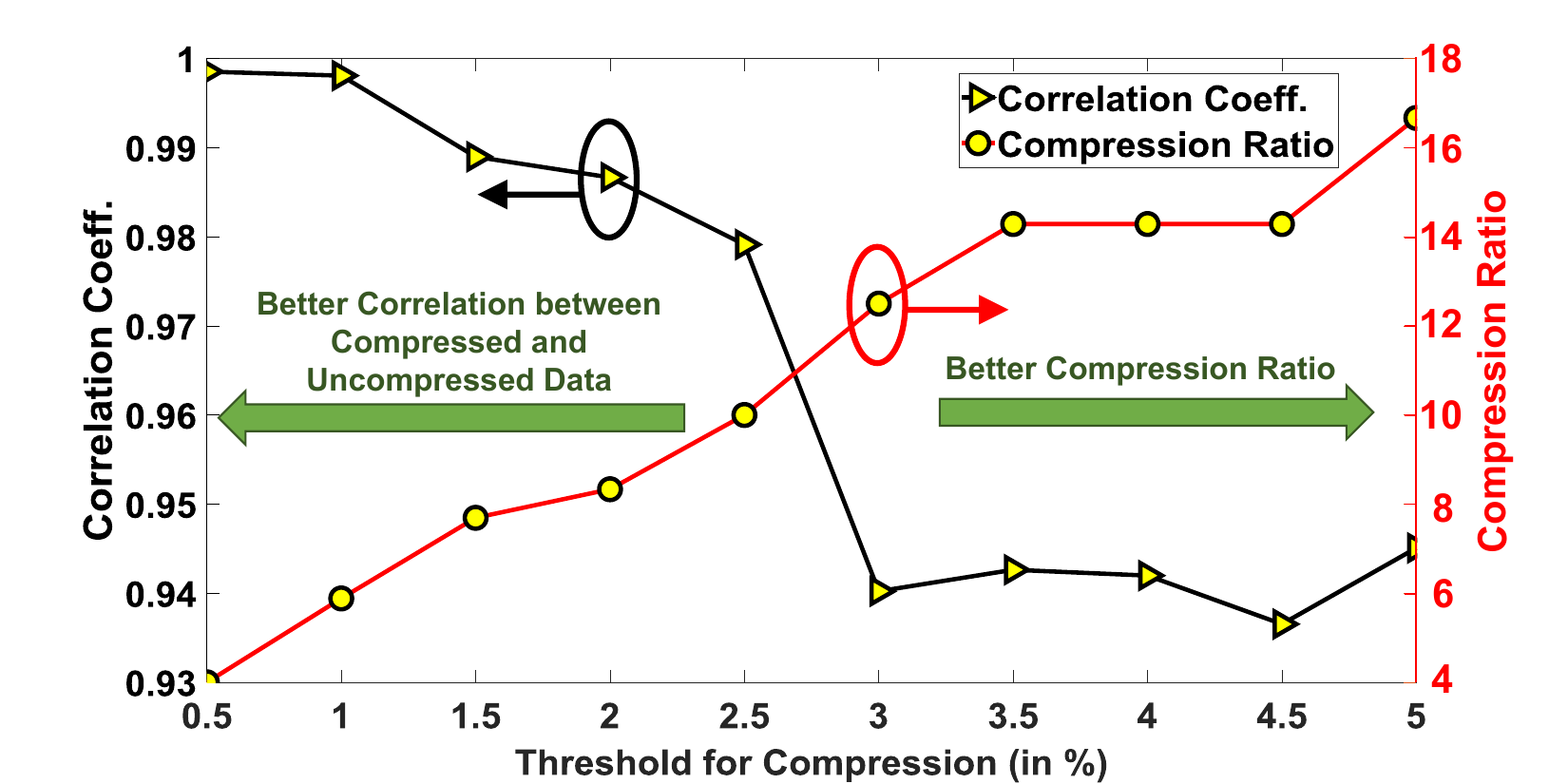}
\caption{Measured Correlation coefficient (left y-axis) and Compression ratio (right y-axis) as the threshold for compression is varied from 0.5\% to 5\% with the temperature data presented in Fig \ref{compress1}.}
\label{compress2}
\end{figure}

Fig. \ref{compress2} shows the trade-off between the measured compression ratio and the correlation coefficient between uncompressed and compressed data, as the threshold for compression ($y\%$) is varied from 0.5\% to 5\%. At lower thresholds, a better correlation is observed between the uncompressed and compressed data ($>$0.98 for $y<2$), while for higher thresholds ($y>3$), better compression ratios ($>12$X) are achieved, still maintaining a correlation coefficient $>$ 0.9. Based on this analysis, a threshold of 2\% was chosen for compression, resulting in a correlation coefficient of $>$0.98. 

\begin{figure}[!t]
\centering
\includegraphics[width=3in]{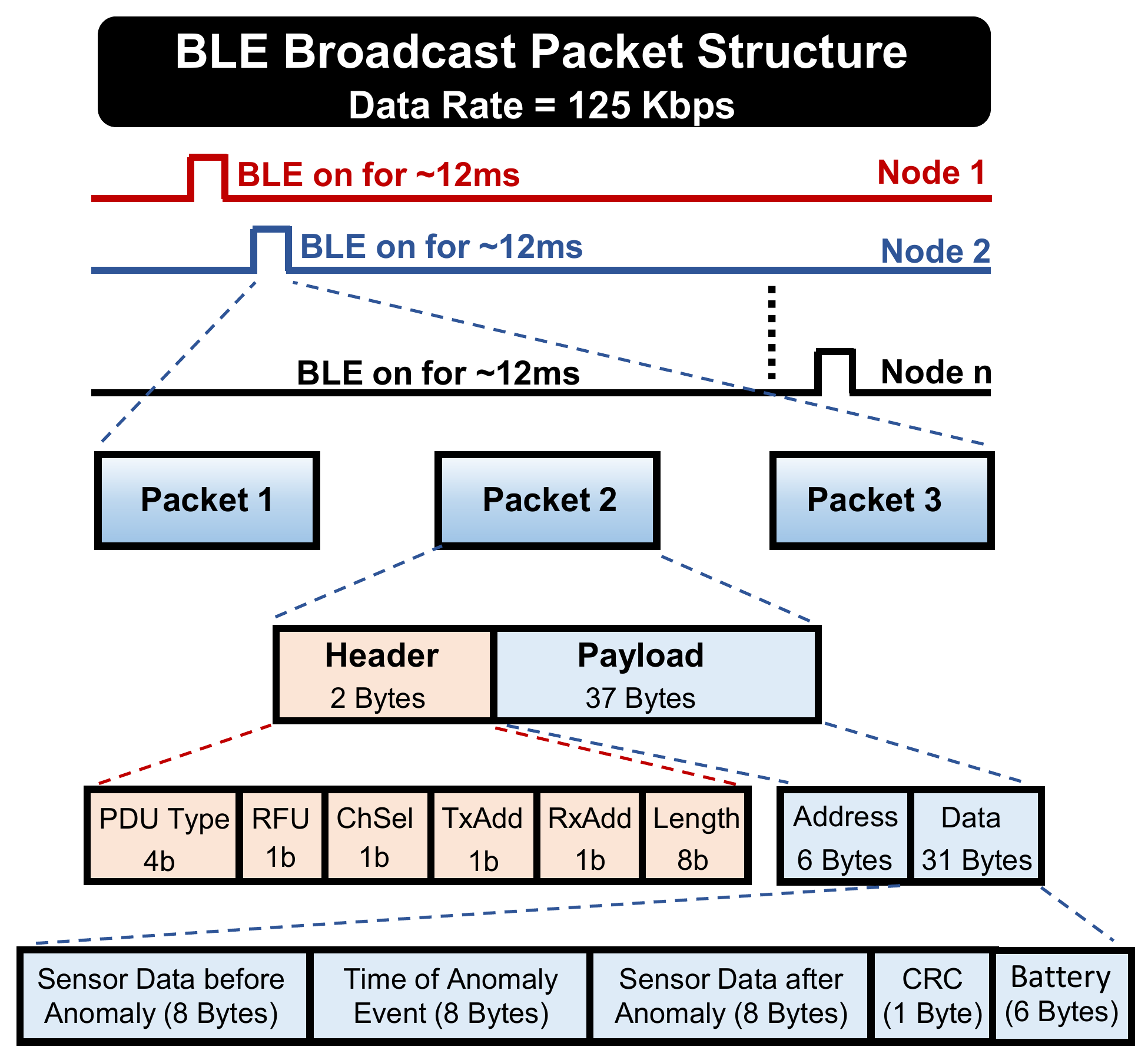}
\caption{BLE broadcast packet structure for each slave device at 125 kbps for short range communication. BLE operation is time-multiplexed for the each node to avoid conflicts. Every BLE broadcast event cosists 3 packets with 39 Bytes each. Each packet contains 31 Bytes processed data from a particular sensor (temperature, humidity and nitrate concentration), which contains information on the sensor data before the occurrence of anomaly (8 Bytes), time of anomaly (8 Bytes), sensor data after the occurrence of anomaly (8 Bytes), and the remaining battery information for the node (6 Bytes). Details on each field within the packet Header and Payload can be found in \cite{BLESpec}.}
\label{ci_ble}
\end{figure}

\subsection{Short Range BLE communication for CI and CAS}
Fig. \ref{ci_ble} shows the BLE communication packet structure for 125 kbps short range data transfer, which is implemented in broadcast mode. 3 packets of 39 Bytes each are transmitted from each BLE slave node when an anomaly occurs, with a 12ms connection time. The transmission from different slaves are time-multiplexed by hard-coding a predefined delay in each node which determines the amount of time after which BLE communication takes place following an anomaly event (for that particular node). The master node (cluster-head for CAS and LoRa transmission) can connect to 7 slave nodes at once due to BLE pairing limitations \cite{BLErev}, and is initially selected by the network designer during initialization. During normal operation, the role of the cluster-head is dynamically updated based on the highest available battery life.

As shown in Fig. \ref{ci_ble}, each packet contains 31 bytes of data payload, out of which 8 Bytes are used to represent the sensor data before the occurrence of an anomaly, 8 Bytes are for time of anomaly and 8 Bytes are used to represent the sensor data after the anomaly. 6 Bytes are used to transmit information on the available battery life. Each device also sends its specific software defined Device ID. The master maintains arrays of anomaly information (data before the anomaly, time of anomaly and data after the anomaly) along with remaining battery life for each node, and performs long-distance LoRa operation for a cluster. At the network level, the clusters are formed based on the algorithm presented in \cite{cluster}. The master can also assign the task of LoRa transmission to other nodes, if more than one cluster of spatially compressed data is found. Once the battery-level of the master drops below any of the slave's battery (due to continuously running CI and CAS algorithms and performing event-driven LoRa transmissions), it broadcasts a message with the corresponding device ID that would become the next master, and switches itself to a slave device once the new master becomes operational. For long-range, a multi-hop LoRa mesh network is implemented that helps sending data to long distances ($>$ 2.5 km) with $\approx$270 mW power consumption (LoRa Tx+Rx) at 10 dBm power output. The network-level details of the short-range and long-rage communication is out of scope of this paper. 


\section{System Level Measurement Results}
\label{measured}

\begin{figure}[!t]
\centering
\includegraphics[width=3.5in]{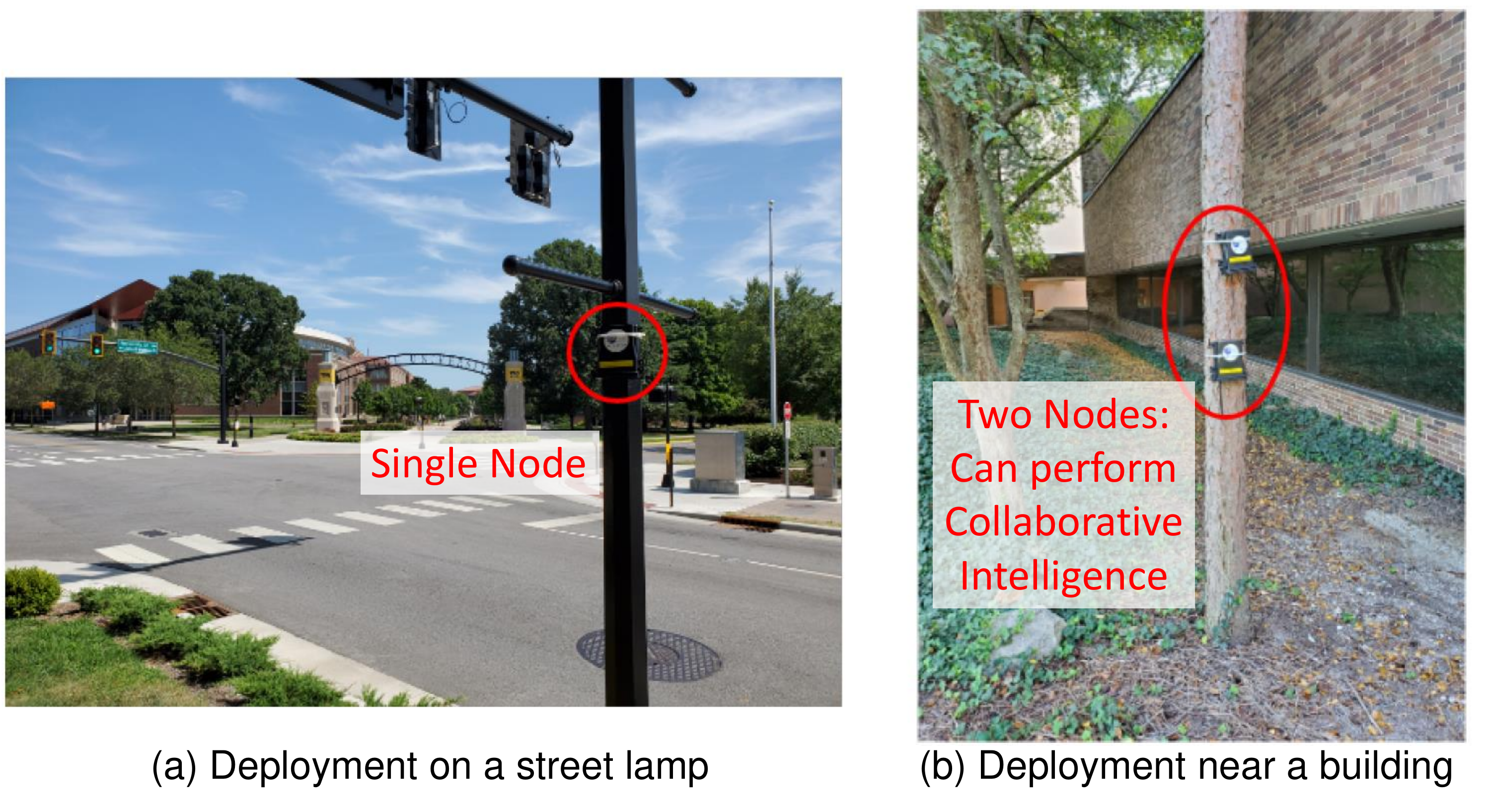}
\caption{Examples of Deployment of the sensor nodes around Purdue campus (a) On a street lamp, (b) Near a campus building.}
\label{street_lamp}
\end{figure}

\begin{figure*}[!t]
\centering
\includegraphics[width=6in]{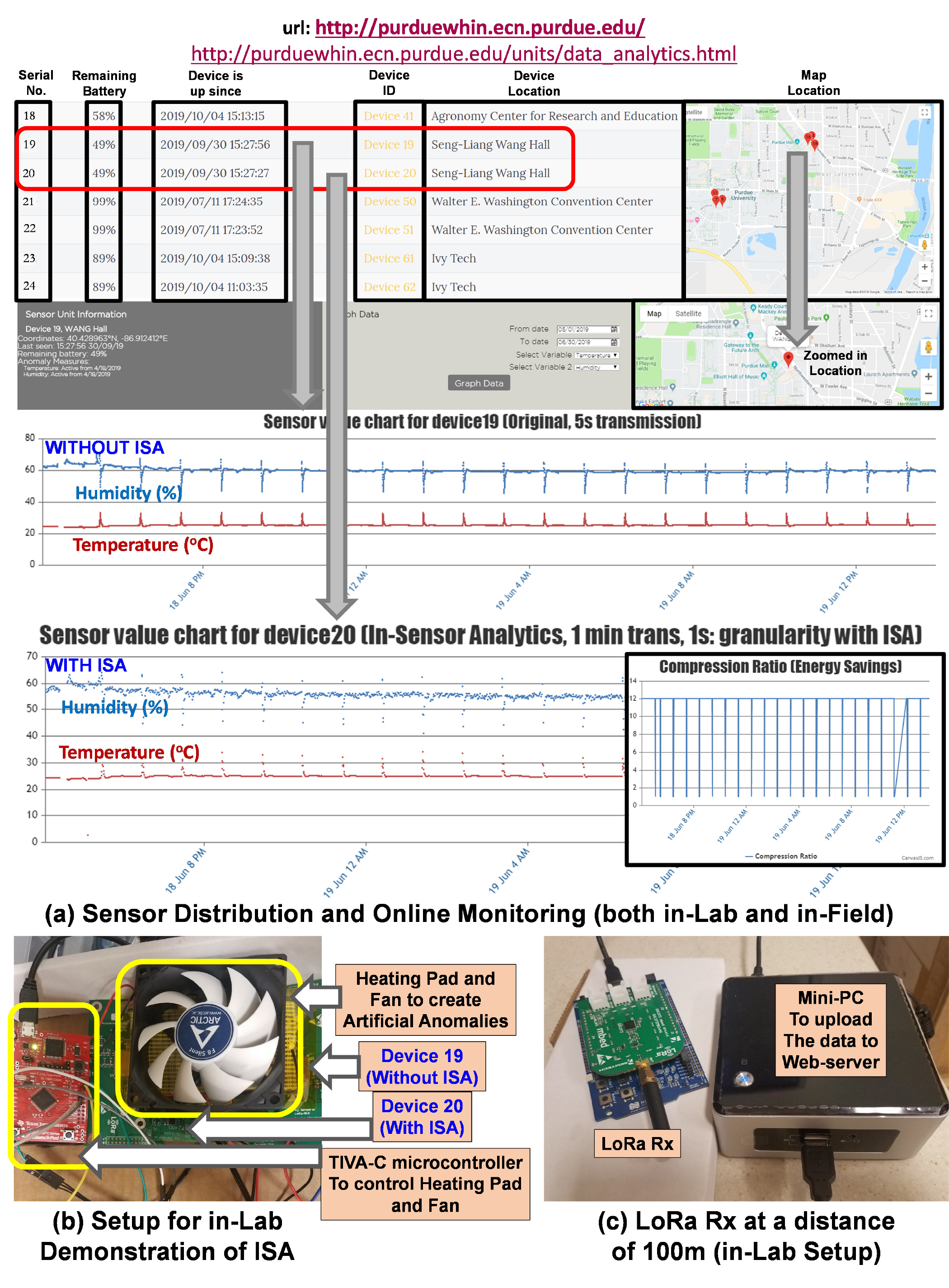}
\caption{(a) Sensor distribution and live monitoring as performed in \url{http://purduewhin.ecn.purdue.edu/units/sensor_list.html}. Device 19 (without ISA) and Device 20 (with ISA) are placed on the same location, $<$ 1 foot away from each other. Both Device 19 and device 20 samples once every 5 seconds. Device 19 transmits the data when it is sampled, while Device 20 compresses the data and sends out when there is an anomaly, which is artificially created every minute using a heating pad and coolig fan for demonstration purpose. This demonstration achieves energy savings of $\approx$12X, simply due to the ratios of transmission time; (b) Setup for in-Lab demonstration of ISA with Device 19 and Device 20; (c) LoRa Rx at a distance of 100 m (in-Lab Setup).}
\label{sensor_dist}
\end{figure*}

\subsection{Deployment of Sensors}
More than 25 sensors are deployed at 5 different geographical locations in and around the 2400-acre Purdue University campus, West Lafayette, Indiana, USA, with permissions from the university's Department of Environmental Health and Public Safety. Fig. \ref{street_lamp} shows two examples of deployed sensors around the Purdue campus. A web-based framework for live monitoring of the sensors is developed (\url{http://purduewhin.ecn.purdue.edu/units/sensor_list.html}) which is shown in Fig. \ref{sensor_dist}. The web-based platform shows the information about the devices (battery life, up-time, location etc.) and graphically plots the read-out data (for example, temperature and humidity) as time-varying quantities. Fig. \ref{sensor_dist} also shows data from two devices (device 19 - without ISA and device 20 - with ISA) which are placed in the same location, $<$ 1 foot away from each other. For demonstration purpose, temperature anomalies are created artificially using a heating pad and a cooling fan every 60 seconds. Device 20 sends out the compressed data at the occurrence of an anomaly, while device 19 sends out uncompressed data once every 5 seconds. This demonstration achieves an energy saving of about 12X, due to the ratios of the transmission interval of device 20 and device 19.

\subsection{Analysis of Energy Consumption}

\begin{figure}[!t]
\centering
\includegraphics[width=3.5in]{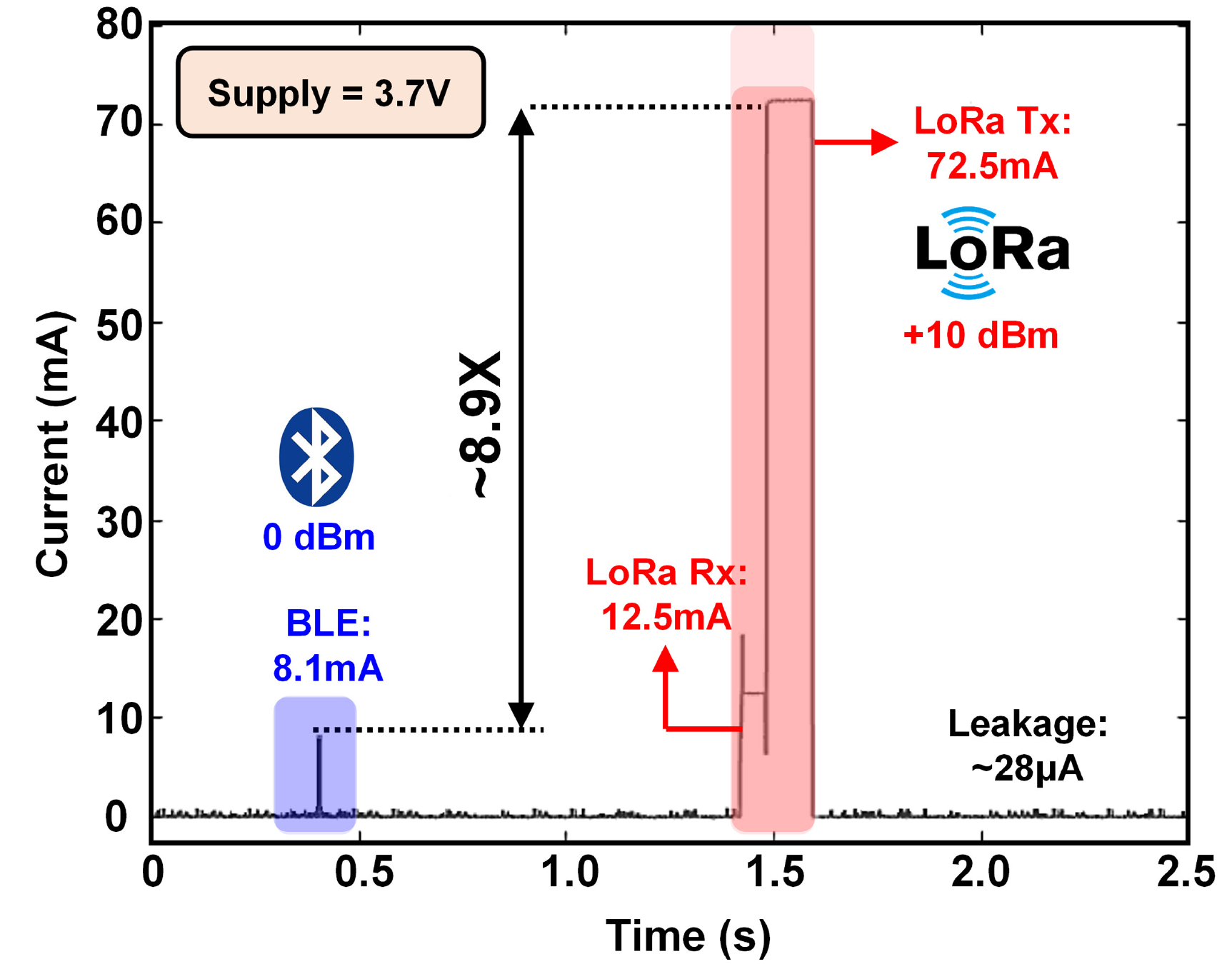}
\caption{Measured Current Consumption during one transmission Cycle. The blue region represents BLE communication window, while the red region represents LoRa communication window.}
\label{current_ble_LoRa}
\end{figure}

\begin{figure}[!t]
\centering
\includegraphics[width=3.5in]{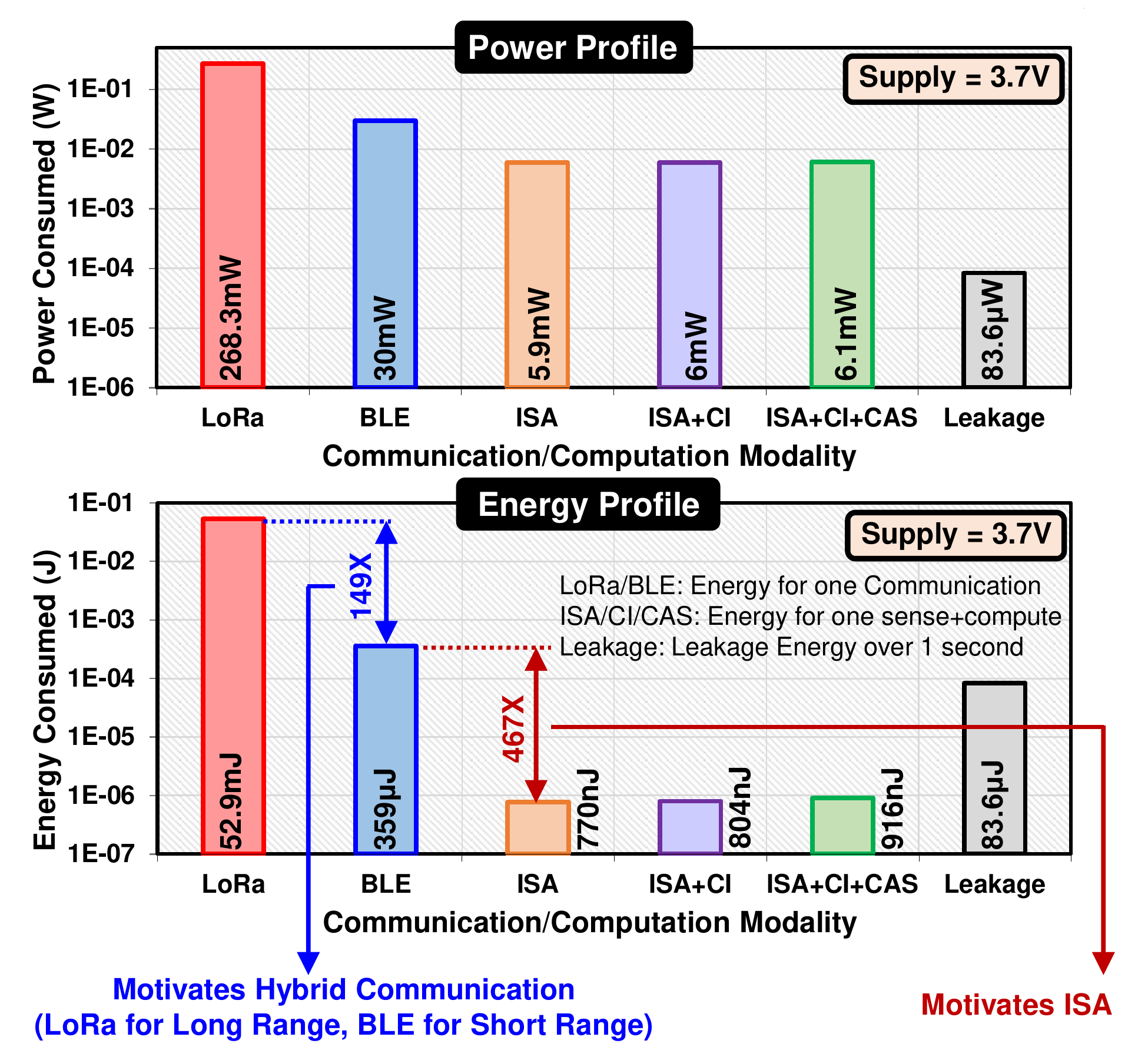}
\caption{Measured Power and Energy Profiles from the nRF52-based sensor node. For communication energy calculations, a single event is considered (LoRa: 72.5 mA transmission over ~130ms, 12.5 mA reception over ~60 ms, BLE: 8.1 mA over ~12 ms). Similarly, one sense+compute event is considered for ISA (3.6 mA for ~130 $\mu$s), ISA+CI (3.61 mA for ~135 $\mu$s) and ISA+CI+CAS (3.65 mA for ~150 $\mu$s). The leakage current is almost constant at ~28 $\mu$A.}
\label{energy_profile}
\end{figure}

\begin{figure}[!t]
\centering
\includegraphics[width=3.5in]{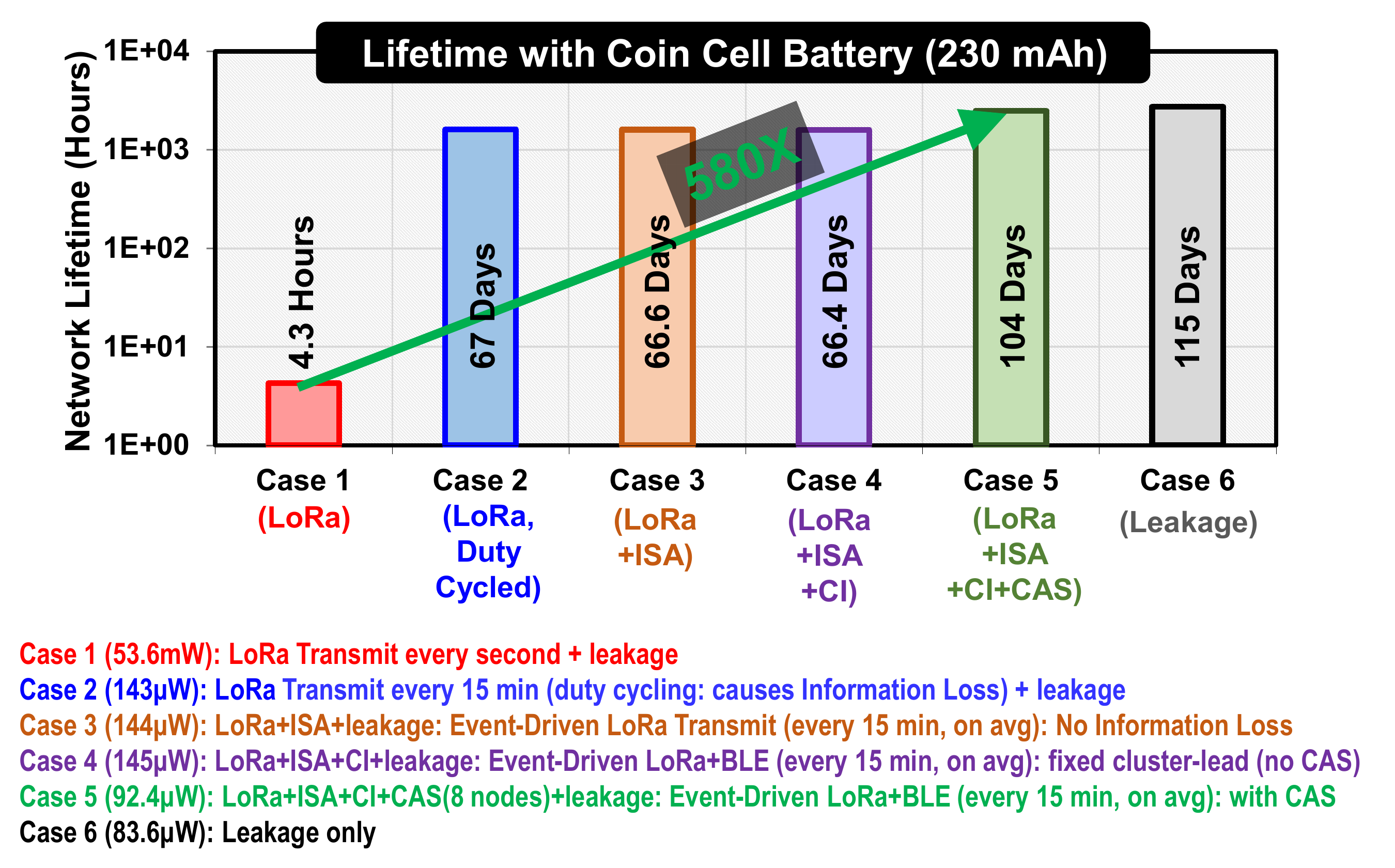}
\caption{Estimated Network Lifetime for different scenarios presented in this paper, including non-duty cycled and duty-cycled LoRa, ISA, ISA+CI and ISA+CI+CAS.}
\label{nlife}
\end{figure}

The measured current consumption profiles for BLE and LoRa are shown in Fig. \ref{current_ble_LoRa}. BLE consumes a maximum current of 8.1 mA at 3.7 V supply for 0 dBm transmission, while the LoRa Tx consumes 72.5 mA for 10 dBm transmission, also at 3.7 V supply. The Rx in LoRa also consumes 12.5 mA for multi-hop communication. The power and energy profiles of the entire nRF52-based sensor node is presented in Fig. \ref{energy_profile} for different communication modalities (LoRa/BLE) and for different concepts presented in this paper (ISA, ISA+CI and ISA+CI+CAS). The 8.9X gap between BLE and LoRa currents and the shorter duration of BLE communication (12ms) as compared to LoRa (Rx+Tx for a total of 190ms) supports our approach of using BLE for short-range communication. For calculation of communication energies, a single event is considered, which can be either (1) a 12.5 mA LoRa reception over 60 ms followed by a 72.5 mA LoRa transmission over 130 ms, or (2) a 8.1 mA BLE communication over 12 ms. In a similar manner, one sense and compute event is considered for calculating the energies for ISA (3.6 mA for 130 $\mu$s), ISA+CI (3.61mA for 135$\mu$s) or ISA+CI+CAS (3.65 mA for 150 $\mu$s). In terms of overall energy consumption, BLE is 149X lower than LoRa (which motivates the need for hybrid communication in the node), while ISA consumes $\approx$467X lower power than BLE, which motivates trading-off computation energy to save more communication energy. ISA+CI+CAS consumes $\approx$282X lower power than BLE while helping to improve the network lifetime, as would be shown in Fig. \ref{nlife}. It is also interesting to note in Fig. \ref{energy_profile} that ISA+CI+CAS consumes lower energy (916 nJ) than the leakage (83.6 $\mu$J) over 1 second, which means that by proper application of the ISA+CI+CAS algorithms along with event-driven communication, it is possible to achieve high network lifetime, primarily limited by the leakage current in the sensor node. A constant leakage/sleep current of $\approx$28 $\mu$A was measured in the nRF52-based custom-developed board, when no communication and/or computation takes place.

Fig. \ref{nlife} presents an estimate of the battery-life of the sensor node, with a 230 mAh coin cell battery, and for the various cases presented in the paper. If we only transmit using LoRa once every second, the battery is drained out in a mere 4.3 hours. With duty cycled LoRa (one transmission every 15 minutes), the battery life increases to 67 days. However, this causes loss of information. The ISA algorithms aim at avoiding this loss of information ($>$98\% retention of data) with a similar network lifetime, if we assume that the average rate of anomaly is still once every 15 minutes. In fact, both ISA and ISA+CI algorithms would reduce the network lifetime by a small amount as the fixed master node consumes more energy (and becomes the bottleneck in calculating network lifetime) if the rate of LoRa transmission is kept constant. However, when we enable ISA+CI+CAS, the network lifetime increases to 104 days, as the burden of LoRa transmission is amortized over dynamic master nodes in the network. This is noticeably close to the theoretical limit of 115 days as posed by the leakage currents in the nodes, and is a 580X improvement over the case with 1 LoRa transmission per second, while maintaining $>$98\% of information. 


\section{Conclusion}
\label{conclusion}

In conclusion, this paper extensively analyzes the communication-computation trade-offs in a wireless sensor network, and proposes a combination of energy-optimization strategies, both at the sensor and network level, in the form of (1) In-Sensor Analytics (ISA) that enables event driven communication (through anomaly detection) and temporal data compression, (2) Collaborative Intelligence (CI) that makes use of short-range BLE for spatial data compression, (3) Context-Aware Switching (CAS) based on available battery-life for network longevity, by enabling only the master/cluster-head in CI to perform long-distance LoRa communication, (4) multi-hop LoRa which allows to have a longer range with a lower communication energy. The proposed computation algorithms (ISA+CI+CAS) consume $\approx$282X lower energies than BLE and  $\approx$57750X lower energies than LoRa, while allowing effective transmission information sampled every second. This helps us to increase the overall network lifetime to the levels almost limited by the system's leakage power (a 580X improvement over LoRa transmissions every second) with $\approx$98\% correlation with the uncompressed data. A web-based monitoring framework is also developed to continuously archive the data and register anomalies, along with a provision to demonstrate the energy-benefits obtained from the proposed ISA in an online manner. As a future work, analysis of energy consumption would be performed throughout the year for reliability over different weather conditions. Solar/RF energy harvesting techniques (instead of using coin cell batteries) would be adopted for continuous operation, and the security of the sensor devices would be analyzed \cite{RF-PUF}. Additionally, since the application involves slowly-varying signals such as temperature, humidity and water nitrate concentration, ultra-low power sensor electronics (developed in-house \cite{JSSC_Rad}) that utilizes the slow nature of the signals could be integrated with the sensor nodes in future.




\ifCLASSOPTIONcaptionsoff
  \newpage
\fi


\bibliographystyle{ieeetr}
\small
\bibliography{bib_file}


%

\begin{IEEEbiography}[{\includegraphics[width=1in,height=1.25in,clip,keepaspectratio]{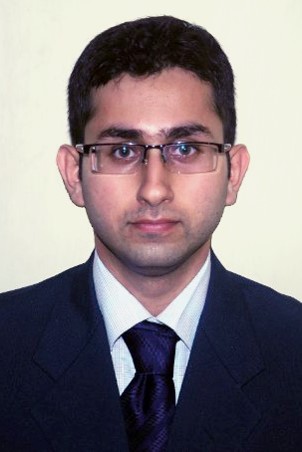}}]
{Baibhab Chatterjee}
 (S`17) received the B.Tech. degree in electronics and communication engineering from the National Institute of Technology (NIT), Durgapur, India, in 2011, and the M.Tech. degree in electrical engineering from IIT Bombay, Mumbai, India, in 2015. He is currently pursuing the Ph.D. degree with the School of Electrical Engineering, Purdue University, West Lafayette, IN, USA. His industry experience includes two years as a Digital Design Engineer/a Senior Digital Design Engineer with Intel, Bengaluru, India, and one year as a Research and Development Engineer with Tejas Networks, Bengaluru. His research interests include low-power analog, RF, and mixed-signal circuit design for secure biomedical applications.
 
 Mr. Chatterjee received the University Gold Medal from NIT, Durgapur, India, in 2011, the Institute Silver Medal from IIT Bombay in 2015, the Andrews Fellowship at Purdue University during 2017-2019, the HOST 2018 Best Student Poster Award (3rd), the HOST 2019 Best Student Paper Award (co-authored) and the CICC 2019 Best Paper Award (overall). 
\end{IEEEbiography}

\vspace{1mm}
\begin{IEEEbiography}[{\includegraphics[width=1in,height=1.25in,clip,keepaspectratio]{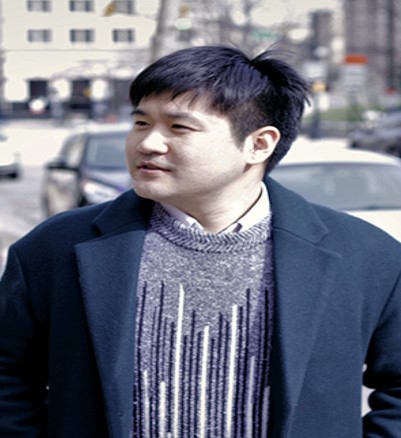}}]
{Donghyun Seo}
received the B.S. degree in electronics and radio engineering from Kyung Hee University, Seoul, South Korea, in 2013, and the M.S. degree in electronics computer engineering at Hanyang University, Seoul, South Korea, in 2015, and is currently working toward Ph.D. degree in the School of Electrical Engineering, Purdue University, West Lafayette, IN, USA. His research interests include CMOS low-power analog, mixed signal and RF integrated circuit design for sensor node interfacing.
\end{IEEEbiography}

\vspace{1mm}
\begin{IEEEbiography}[{\includegraphics[width=1in,height=1.25in,clip,keepaspectratio]{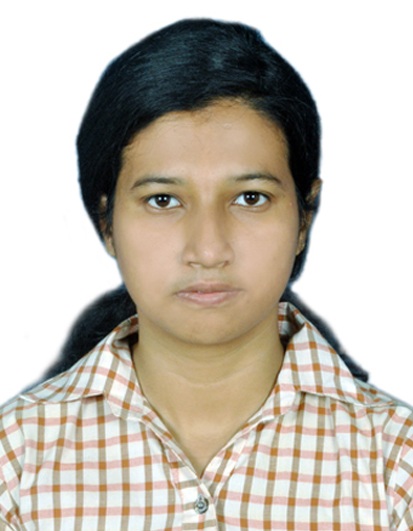}}]
{Shramana Chakraborty}
received the B.Tech degree in Electronics and Communication Engineering from Maulana Abul Kalam Azad University of Technology (MAKAUT, Formerly WBUT), Kolkata, India in 2016 and has recently received M.S. degree in Electrical and Computer Engineering from Purdue University, West Lafayette, IN, USA in 2020. She is the recipient of University Gold Medal from MAKAUT, Kolkata in 2016. Her area of interests include implementation of embedded systems, hardware-software interfacing, and computer networking.
\end{IEEEbiography}

\vspace{1mm}
\begin{IEEEbiography}[{\includegraphics[width=1in,height=1.25in,clip,keepaspectratio]{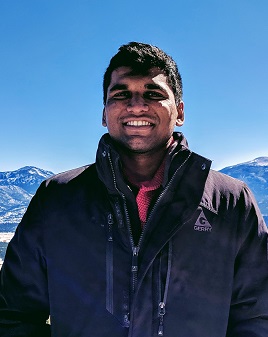}}]
{Shitij Avlani}
received his Bachelor of Engineering in Electronics Engineering from the University of Mumbai, India in 2017, during which time he worked at IIT-Bombay for three years on point of care medical devices and fabrication of MEMS sensors. He is currently working towards a M.S degree in the School of Electrical Engineering at Purdue University, West Lafayette, IN, USA. His research interests include embedded systems and analog and mixed signal circuits for sensor nodes.
\end{IEEEbiography}

\vspace{1mm}
\begin{IEEEbiography}[{\includegraphics[width=1in,height=1.25in,clip,keepaspectratio]{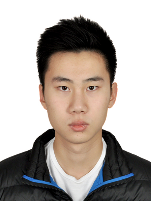}}]
{Xiaofan Jiang}
received the Bachelor's degree in electrical engineering in 2015 from Purdue University, West Lafayette, IN, USA, where he is currently a Ph.D. candidate working in LPWAN technologies for IoT, involving hybrid mesh networks of LoRa, BLE and ANT. He is interested in design and implementation of embedded systems, wireless sensor networks and IoT devices for general and industry applications.

\end{IEEEbiography}

\vspace{1mm}
\begin{IEEEbiography}[{\includegraphics[width=1in,height=1.25in,clip,keepaspectratio]{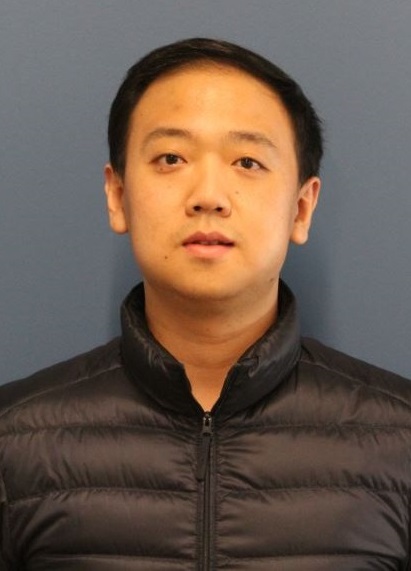}}]
{Heng Zhang}
received his BS degree from Shanghai Jiao Tong Univerisity in 2016 majoring in Electrical and Computer Engineering. He is currently a Ph.D. student working with Prof. Saurabh Bagchi in the Dependable Computing Systems Laboratory at the Department of Electrical and Computer Engineering, Purdue University, West Lafayette, IN, USA. His research focuses on Mobile Crowd-Sensing (MCS), Mobile Computing and Software Defined Network (SDN). Primary techniques used in my work include Network Simulation, System Optimization and Machine Learning.
\end{IEEEbiography}

\vspace{1mm}
\begin{IEEEbiography}[{\includegraphics[width=1in,height=1.25in,clip,keepaspectratio]{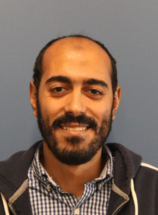}}]
{Mustafa~Abdallah} received the B.Sc. degree in Electronics and Communications Engineering in 2012 and the M.Sc. degree in Engineering Mathematics in 2016, from Faculty of Engineering, Cairo University. He is currently pursuing the Ph.D. degree in Electrical and Computer Engineering at Purdue University, West Lafayette, IN, USA. His research interests are in game theory, behavioral decision-making, and deep learning, with applications including cyber physical systems and speech recognition. He was the recipient of the best fresher award in DCSL lab at Purdue University in 2017 and a M.Sc. fellowship from the Faculty of Engineering, Cairo University in 2013.
\end{IEEEbiography}

\vspace{1mm}
\begin{IEEEbiography}[{\includegraphics[width=1in,height=1.25in,clip,keepaspectratio]{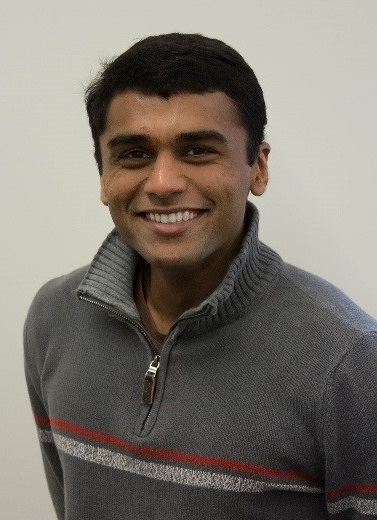}}]
{Nithin Raghunathan}
received the Bachelor's degree in electrical engineering in 2007 and the Ph.D. degree in electrical engineering in 2014, both from Purdue University, West Lafayette, IN, USA, where he is currently working as a research scientist at the Birck Nanotechnology Center. His research interests include novel MEMS devices, sensors and micromachined high-g MEMS accelerometers for impact applications, small-form factor agricultural sensors and ultra-low power wireless platforms for multi-sensor applications.
\end{IEEEbiography}

\vspace{1mm}
\begin{IEEEbiography}[{\includegraphics[width=1in,height=1.25in,clip,keepaspectratio]{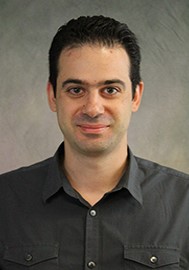}}]
{Charilaos Mousoulis}
(M`11) received the Diploma degree in Electrical and Computer Engineering from the National Technical University of Athens, Athens, Greece, in 2005 and the Ph.D. degree in Electrical Engineering from Purdue University, West Lafayette, IN, USA in 2012. He was a Post-doctoral Research Associate at the School of Biomedical Engineering, Purdue University, West Lafayette during 2011-2014. Since 2014, he is a Senior Research Scientist at the School of Electrical Engineering, Purdue University, West Lafayette. He has extensive experience on the design of silicon-based radiation sensors for occupational dosimetry. His research interests include microsystems for biomedical applications, sensors for food safety, flexible hybrid electronics, and IoT-based sensors for precision agriculture and advanced manufacturing.
\end{IEEEbiography}

\vspace{1mm}
\begin{IEEEbiography}[{\includegraphics[width=1in,height=1.25in,clip,keepaspectratio]{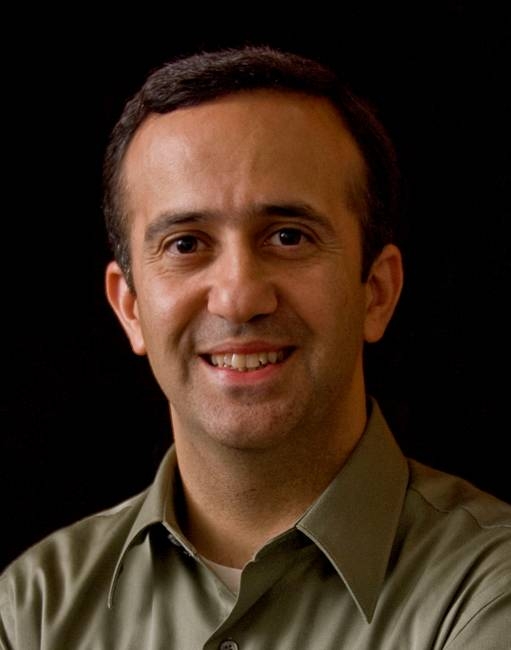}}]
{Ali Shakouri}
(S`xx-M`01-SM`xx-F`xx) received the Bachelor.s degree in engineering from Telecom ParisTech, Paris, France, in 1990 and the Doctoral degree from the California Institute of Technology, Pasadena, CA, USA, in 1995. He is currently the Mary Jo and Robert L. Kirk Director of the Birck Nanotechnology Center and a Professor with the Department of Electrical and Computer Engineering, Purdue University, West Lafayette, IN, USA. From 1998 to 2011, he was a faculty member with the University of California, Santa Cruz, where he directed a multi university research center focused on direct conversion of heat into electricity. He also initiated a sustainability curriculum and a California-Denmark summer program in renewable energies in collaboration with colleagues in sociology, political science, and environmental studies. His major initiative at the Birck Center focuses on nanomanufacturing and printing smart films. The goal is to develop low-cost devices and Internet of Things sensor networks to enable digital agriculture, smart food packaging, wearables for healthcare monitoring, and smart infrastructure. This involves two dozen faculty from colleges of engineering, science, agriculture, and pharmacy. He is a recipient of a Packard Fellowship in Science and Engineering in 1999, an NSF CAREER Award in 2000, and the Thermi Award in 2014.
\end{IEEEbiography}

\vspace{1mm}
\begin{IEEEbiography}[{\includegraphics[width=1in,height=1.25in,clip,keepaspectratio]{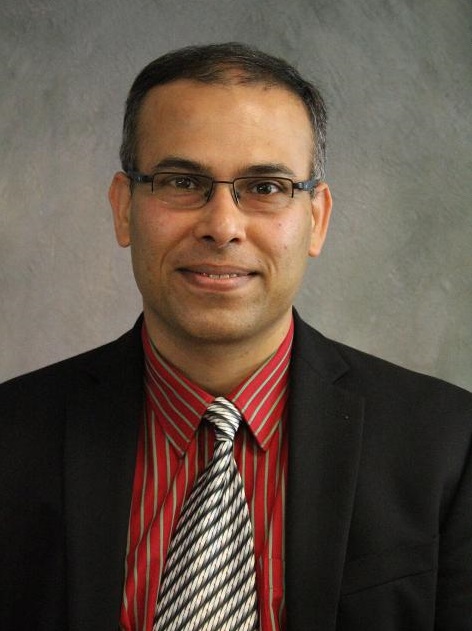}}]
{Saurabh Bagchi}
(S`xx-M`01-SM`xx) is a Professor in the School of Electrical and Computer Engineering and in the Department of Computer Science at Purdue University, West Lafayette, Indiana. He is an ACM Distinguished Scientist and a senior member of IEEE and ACM. At Purdue, he is the Assistant Director of CERIAS, the security center, an IMPACT Faculty Fellow, and leads the Dependable Computing Systems Laboratory (DCSL) where his group performs research in practical system design and implementation of dependable distributed systems. He has been a Visiting Scientist with IBM's Austin Research Lab since 2011.
\end{IEEEbiography}

\vspace{1mm}
\begin{IEEEbiography}[{\includegraphics[width=1in,height=1.25in,clip,keepaspectratio]{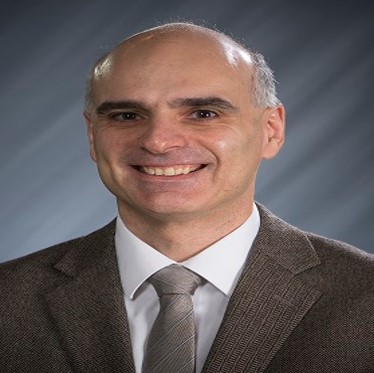}}]
{Dimitrios Peroulis}
(S`99-M`04-SM`15-F`17) received the Ph.D. degree in electrical engineering from the University of Michigan, Ann Arbor, MI, USA, in 2003. He is currently a Professor of Electrical and Computer engineering, the Deputy Director of the Birck Nanotechnology Center and the Michael and Katherine Birck Head of the School of Electrical and Computer Engineering at Purdue University. He has co-authored over 300 journal and conference papers. His current research interests are focused on the areas of reconfigurable electronics, cold-plasma RF electronics, and wireless sensors. He has been a key contributor on developing high quality reconfigurable filters and filter synthesis techniques based on tunable miniaturized high-Q
resonators. He is also leading unique research efforts in high-power multifunctional RF electronics. He received the National Science Foundation CAREER Award in 2008. In 2014 he received the Outstanding Young
Engineer Award of the IEEE Microwave Theory and Techniques Society. In 2012 he received the Outstanding Paper Award from the IEEE Ultrasonics, Ferroelectrics, and Frequency Control Society (Ferroelectrics section). His students have received numerous student paper awards and other student research-based scholarships. He has been a Purdue University Faculty Scholar and has also received ten teaching awards, including the 2010 HKN C. Holmes MacDonald Outstanding Teaching Award and the 2010 Charles B. Murphy Award, which is Purdue University`s highest undergraduate teaching honor.
\end{IEEEbiography}

\vspace{1mm}
\begin{IEEEbiography}[{\includegraphics[width=1in,height=1.25in,clip,keepaspectratio]{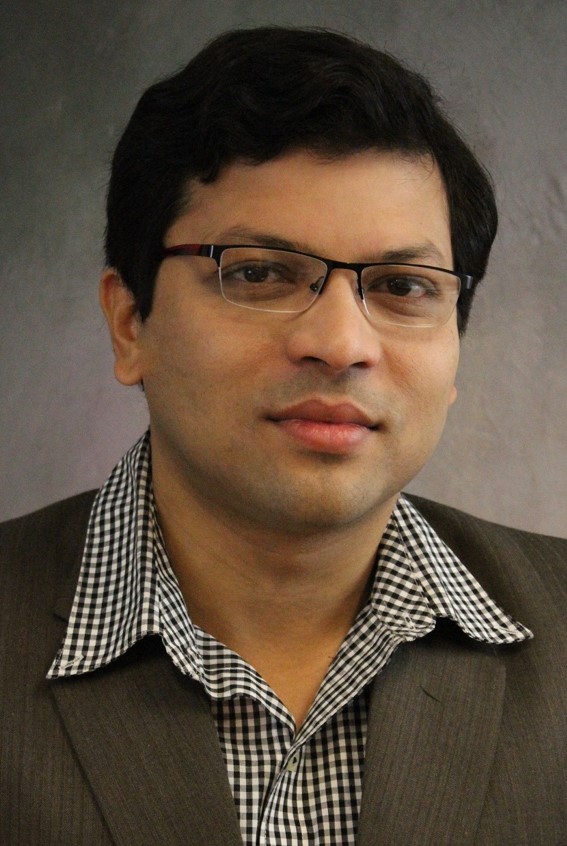}}]
{Shreyas Sen}
(S`06-M`11-SM`17) received the Ph.D. degree in electrical and computer engineering from Georgia Tech, Atlanta, GA, USA, in 2011. He has over five years of industry research experience with Intel Labs, Hillsboro, OR, USA, Qualcomm, Austin, TX, USA, and Rambus, Los Altos, CA, USA. He is currently an Assistant Professor with the School of Electrical and Computer Engineering, Purdue University, West Lafayette, IN, USA. He has authored or coauthored two book chapters and over 130 conference and journal articles. He holds 14 patents granted/pending. His research interests include mixed signal circuits/systems for the Internet of Things (IoT), biomedical, and security.

Dr. Sen was chosen by MIT Technology Review as one of the top ten Indian Inventors Worldwide under 35 (MIT TR35 India Award), in 2018, for the invention of using the human body as a wire, which has the potential to transform healthcare, neuroscience, and human-computer interaction. His work has been covered by more than 100 news releases worldwide, invited appearance on TEDx Indianapolis, Indian National TV CNBC TV18 Young Turks Program, and by Radio Interview on NPR subsidiary Lakeshore Public Radio. He was a recipient of the AFOSR Young Investigator Award in 2017, the NSF CISE CRII Award in 2017, the Google Faculty Research Award in 2017, the HKN Outstanding Professor Award, the Intel Labs Divisional Recognition Award in 2014, for industry-wide impact on USB-C type, the Intel Ph.D. Fellowship in 2010, the IEEE Microwave Fellowship in 2008, the GSRC Margarida Jacome Best Research Award in 2007, the Best Paper Awards at CICC 2019 and HOST 2017, 2018, and 2019, the ICCAD Best-in-Track Award in 2014, the VTS Honorable Mention Award in 2014, the RWS Best Paper Award in 2008, the Intel Labs Quality Award in 2012, the SRC Inventor Recognition Award in 2008, and the Young Engineering Fellowship in 2005. He serves\/has served as an Associate Editor for the IEEE DESIGN AND TEST, an Executive Committee Member of the IEEE Central Indiana Section, ETS, and a Technical Program Committee Member of DAC, CICC, DATE, ISLPED, ICCAD, ITC, VLSI Design, IMSTW, and VDAT.
\end{IEEEbiography}






\end{document}